\documentclass[12pt,preprint,a4paper]{aastex}
\usepackage{times}
\usepackage{graphics,epsf}
\usepackage{amsmath}                
\usepackage{amsfonts}               
\usepackage{amssymb}                
\usepackage{epsfig}                 
\usepackage{graphicx}
\usepackage{url}
\usepackage{tabularx}
\usepackage{booktabs}
\usepackage{epstopdf}
\usepackage{subfigure}

\def \cm{~\rm{cm}}
\def \s{~\rm{s}}
\def \km{~\rm{km}}

\def \K{~\rm{K}}
\def \g{~\rm{g}}

\def \AU{~\rm{AU}}

\def \yrs{~\rm{yrs}}
\def \yr{~\rm{yr}}

\def \kpc{~\rm{kpc}}

\begin{document}
\title{The formation of `columns crowns' by jets interacting with a circumstellar dense shell}

\author{Muhammad Akashi\altaffilmark{1,2}, and Noam Soker\altaffilmark{1,3}}

\altaffiltext{1}{Department of Physics, Technion -- Israel
Institute of Technology, Haifa 32000, Israel; akashi@physics.technion.ac.il; 
soker@physics.technion.ac.il}
\altaffiltext{2}{Kinneret College on the Sea of Galilee, Israel}
\altaffiltext{3}{Guangdong Technion Israel Institute of Technology, Shantou 515069, Guangdong Province, China}

\begin{abstract}
We conduct three-dimensional hydrodynamical simulations of two opposite jets that interact with a spherical slow wind that includes a denser shell embedded within it, and obtain a bipolar nebula where each of the two lobes is composed of two connected bubbles and Rayleigh-Taylor instability tongues that protrude from the outer bubble and form the `columns crown'. The jets are launched for a short time of 17 years and inflate a bipolar nebula inside a slow wind. When the bipolar structure encounters the dense shell, the interaction causes each of the two lobes to split to two connected bubbles. The interaction is prone to Rayleigh-Taylor instabilities that form tongues that protrude as columns from the outer bubble. The bases of the columns form a ring on the surface of the outer bubble, and the structure resemble a crown that we term the columns crown.
This structure resembles, but is not identical to, the many filaments that protrude from the lobes of the bipolar planetary nebula Menzel~3. We discuss our results in comparison to the structure of Menzel~3 and the ways by which the discrepancies can be reconciled, and possibly turn our failure to reproduce the exact structure of Menzel~3 to a success with jets-shell interaction simulations that include more ingredients. 
\end{abstract}


\section{INTRODUCTION}
\label{sec:intro}

The interaction of a binary companion with an evolve giant star, like an asymptotic giant branch (AGB) star that is a progenitor of a planetary nebula (PN), has in principle two sources of gravitational energy that might energize the outflow. The first one is the orbital gravitational energy that the binary system releases as the orbital separation between the core of the giant star and the companion decreases. 
This is likely to energize more the equatorial outflow than the polar outflow, either when the secondary star companion is outside the envelope, e.g., by ejecting mass through the second Lagrangian point (e.g., \citealt{Livioetal1979, MastrodemosMorris1999, Pejchaetal2016, Chenetal2017, Pejchaetal2017}) or during the common envelope evolution (e.g., \citealt{Iaconietal2017b, DeMarcoIzzard2017, Galavizetal2017, Iaconietal2017a, Chamandyetal2018, GarciaSeguraetal2018,  MacLeodetal2018}, and references for older than two years papers therein). 

The second energy source is the gravitational energy that the mass that the secondary more compact star accretes from the envelope of the giant star releases. The secondary star can be inside the envelope of the giant star or outside the envelope. The most efficient process to carry the accretion energy to the outflow is to launch jets from an accretion disk. Such jets influence the outflow along and near the polar directions much more than they affect the equatorial outflow. 

In recent years the notion that in many cases jets shape the outflows from evolved stars, and in particular from AGB stars that evolve to become PNe, has benefited from progress in several directions. First is the realization that jets shape many PNe. Although the suggestions for jet-shaping of some PNe is old (e.g., \citealt{Morris1987, Soker1990AJ}), in the last two decades researchers have realized that jets shape a large fraction of non-spherical PNe (e.g., \citealt{SahaiTrauger1998, Boffinetal2012, HuarteEspinosaetal2012, Balicketal2013, Miszalskietal2013, Tocknelletal2014, Huangetal2016, Sahaietal2016, RechyGarciaetal2016, GarciaSeguraetal2016, Dopitaetal2018}). 

Then is the understanding that in many cases the jets operate in a feedback mechanism (see \citealt{Soker2016Rev} for a review). 
In the negative-feedback part the jets remove mass from the ambient medium which serves as the reservoir of accreted gas, hence reducing accretion rate that is followed by a decrease in jets' power (e.g., \citealt{Sokeretal2013, LopezCamaraetal2018}).
The positive-feedback part comes from the removal of energy and gas from the very inner regions of the accretion disk, just near the accreting star. Consequently, this reduces the pressure in those regions and hence allows for a high accretion rate, even at super-Eddington rates (e.g.,  \citealt{Shiberetal2016, Chamandyetal2018}). 

The third direction of progress has been the finding that binary systems shape most, and probably all, PNe (e.g., \citealt{
Akrasetal2016, Alietal2016, Bondetal2016, Chenetal2016, Chiotellisetal2016, GarciaRojasetal2016, Hillwigetal2016a, Jonesetal2016, Madappattetal2016, Chenetal2017, DeMarcoIzzard2017, Hillwigetal2017,JonesBoffin2017, Sowickaetal2017,
Alleretal2018, Barkeretal2018, Bujarrabaletal2018, Ilkiewiczetal2018, Jones2018, Miszalskietal2018b}, for a sample of papers from the last 3 years; for a different model see \citealt{GarciaSeguraetal2005}).
In some cases there is a direct link between the presence of a binary central star and the presence of jets (e.g., \citealt{Boffinetal2012, Miszalskietal2013, Miszalskietal2018a}), and binary AGB systems and the presence of jets launched by the companion to the AGB star (e.g., \citealt{Thomasetal2013, Gorlovaetal2015, Bollenetal2017, VanWinckel2017}). The finding of binary systems in PNe is relevant to the present study because single AGB stars cannot launch jets for the lack of angular momentum, and the presence of a binary companion is crucial for launching jets in these progenitors of PNe.  

In light of the large number of observations that show that jets shape many PNe and other nebulae around evolved stars, we continue our study of the different morphological structures that jets can form. 
We stress here that by jets we refer to any bipolar outflow from the accretion disk, most likely in a binary system. This bipolar outflow can be two opposite narrow jets, a wide bipolar outflow from the accretion disk, it can be a continuous bipolar outflow or a chain of clumps. To all of these and more, we refer as jets. 
There are many simulations of jet-shaping of PNe and related nebulae (e.g., 
\citealt{LeeSahai2004, Dennisetal2009, Leeetal2009, HuarteEspinosaetal2012, Balicketal2013, Akashietal2015, Balicketal2017, Akashietal2018, Balicketal2018}), but here we concentrate on a particular morphological feature that we term columns crown. 

In the present paper we use the hydrodynamical core FLASH (section \ref{sec:numerical}) to study the formation of columns or filaments that protrude from lobes that jets inflate in the direction of the jets initial velocity. 
We describe the formation of the protruding columns, the columns crown, in section \ref{sec:results}. 
We are motivated by such thin columns that are observed in the PN Menzel~3 (Mz~3; PN~G331.7-01.0; the Ant nebula). In an earlier paper \citep{AkashiSoker2008a} we showed that a jet interacting with the AGB wind can form a lobe with a front lobe as observed in Mz~3.
We now perform a simulation with a different set of initial conditions that lead to the formation of a delicate structure of the columns crown.
The PN MZ~3 itself is a well studied PN (e.g., \citealt{Clyneetal2015, Alemanetal2018}) that we discuss more in section \ref{sec:summary} where we summarize our main findings. 
           
\section{NUMERICAL SET-UP}
 \label{sec:numerical}
 
This study is another one in our exploration of the different morphologies that the interaction of jets with an ambient medium can form, with the goal of accounting for the rich variety of morphologies of PNe and other nebulae around evolved stars. We describe here the initial conditions of the simulation that forms more or less straight columns along the polar directions. 

We assume that the ambient medium is a spherical slow wind surrounded by a dense slow spherical shell that itself is surrounded by an outer slow wind zone. A giant star formed the dense shell in an episode of intensive mass loss rate. A main sequence binary companion that is outside the envelope of the giant star launches the two opposite jets at $\simeq 3000 \yr$ after the intensive mass loss episode that formed the dense shell ended. 
  
We simulate the region far from the binary system, and so we ignore the orbital motion and launch the jets along the symmetry axis. As well, the large distance from the binary system allows us to neglect the gravitational field of the binary system, and so we do not include gravity. In other words, the velocities at which we inject the wind and jets are much larger than the escape speed from the regions we simulate. 

We use version 4.2.2 of the hydrodynamical FLASH code \citep{Fryxell2000} with the unsplit PPM (piecewise-parabolic method) solver to perform our 3D hydrodynamical simulations. FLASH is an adaptive-mesh refinement (AMR) modular code used for solving hydrodynamics and magnetohydrodynamics problems. 
 We include radiative cooling of the optically thin gas, and take the cooling function for solar abundance from \cite{SutherlandDopita1993}. We turn off radiative cooling below a gas temperature of $10^4 \K$. 

We employ a full 3D AMR (7 levels; $2^{10}$ cells in each direction) using a Cartesian grid $(x,y,z)$ with outflow boundary conditions at all boundary surfaces. We take the $z=0$ plane to be in the equatorial plane of the binary system, that is also the equatorial plane of the nebula, and we simulate the whole space (the two sides of the equatorial plane).

At time $t=0$ we place a spherical dense shell in the zone $10^{17} \cm < r <  1.75 \times 10^{17} \cm$ and with a density profile of $\rho_s = 4 \times 10^{-22} (r/10^{17} \cm)^{-2} \g \cm^{-3}$, such that the total mass in the shell is $0.002M_\odot$. 
The gas in the shell has an initial radial velocity of $v_s = 10 \km \s^{-1}$. 
The spherical expanding shell corresponds to a wind with a terminal velocity of $v_s = 10 \km \s^{-1}$ that the giant star blown for about 2400 years at a mass loss rate of $\dot M_s=8 \times 10^{-7} M_\odot \yr^{-1}$.
 The regions outside and inside the dense shell are filled with a much lower density gas, the spherically slow wind, with velocity of $v_{\rm wind}=v_s= 10 \km \s^{-1}$ and a constant mass-loss rate of $\dot M_{\rm wind}= 10^{-7}{\rm M_\odot \yr^{-1}}$.

We note that such a shell embedded inside a slow wind is not observed in any PN (e.g., \citealt{Sahaietal2011}). We explain this by the combination of two properties of the flow. Firstly, the ejection of such a shell due to a binary interaction is rare. Secondly, we expect that the interaction of the jets with the shell takes place closer to the center, namely, during a short time after the ejection of the shell. Most likely, the interaction region will be obscured by dust. 
The shorter distance and time imply higher mass loss rate. For example, we can reduce the radius by a factor of about 20, say, decrease the shell production time to $120 \yr$, and increase the mass loss rate into the shell to $\dot M_s=1.6 \times 10^{-5} M_\odot \yr^{-1}$. The entire dense shell is within $10^{16} \cm$, or $600 \AU$. In that case the life of the dense shell before interaction is only 300 years, and might be hidden by dust at the time of the interaction. After the interaction the structure expands ballistically, to reach the same size as we obtain here. Here we simulate one case to present the principle properties of such an interaction.

We launch the two opposite jets from the inner $10^{16} \cm $ zone along the $z$-axis ( at $x=y=0$) and  within a half opening angle of $\alpha = 25^\circ$. By the term `jets' we refer also to wide outflows, as we simulate here. More generally, we simulate slow-massive-wide (SMW) outflows. 
We terminate the launching of the jets at $t=17\yrs$.
The jets' initial velocity is $v_{\rm jet}=800 \km \s^{-1}$, just a little above the escape speed from a main sequence star. The mass-loss rate into the two jets together is $\dot M_{\rm 2jets} = 2 \times 10^{-6} M_\odot \yr^{-1}$. 
For numerical reasons (to avoid very low densities) we inject a very weak slow wind in the directions where we do not launch the jets, i.e., in the sector $\alpha<\theta<90^\circ$ in each hemisphere (for more numerical details see \citealt{AkashiSoker2013}).

The accretion rate in the disk that launches the jets is about 10 times higher than the mass loss rate in the jets, or $2 \times 10^{-5} M_\odot \yr^{-1}$ in the case we simulate, or $ \approx 10^{-4} M_\odot \yr^{-1}$ had we taken the interaction time to be shorter. This is still below the Eddington accretion rate limit on to a main sequence star. The magnitude of the momentum in the two jets is about $5 \times 10^{36} \g \cm \s^{-1}$. This is much below values that are observed in many bipolar PNe (e.g., \citealt{Bujarrabaletal2018}). The jets' velocity we use is somewhat above the speed of observed collimated outflows in PNe, but it is only a little larger than the escape speed from main sequence stars. As well, the jets' velocity in the proto-PN He~3-1475 is $1200 \km \s^{-1}$ (e.g., \citealt{BorkowskiHarrington2001}). 
We estimate that a jets' velocity of about $400 \km \s^{-1}$ in our simulation, as observed in young stellar objects, would lead to the same qualitative results.  
Overall, we consider the jets' properties we use quite plausible, even if rare. 

The initial temperature of the slow wind, the dense shell, and the jets is $10^4 \K$. The initial jets' temperature has no influence on the results (as long as it is highly supersonic) because the jets rapidly cool due to adiabatic expansion. 

\section{RESULTS}
\label{sec:results}
\subsection{The formation of the bipolar structure}
\label{subsec:bipolar}

\begin{figure}
 \begin{center}
  \hskip -0.9 cm
 \includegraphics[trim= 1.0cm 0.2cm 5.5cm 0.0cm,clip=true,width=0.33\textwidth]{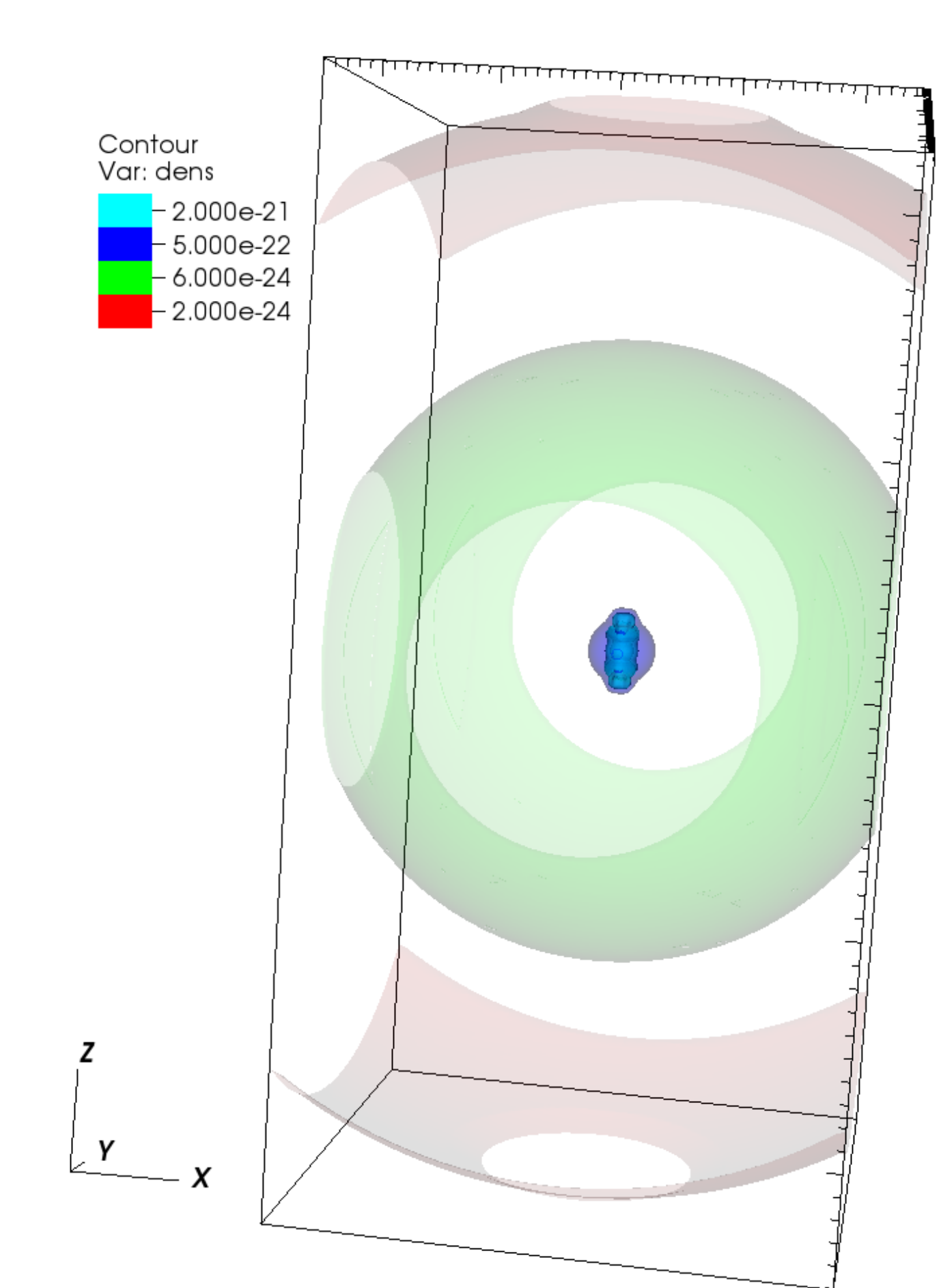}
  \includegraphics[trim= 1.0cm 0.2cm 5.5cm 0.0cm,clip=true,width=0.33\textwidth]{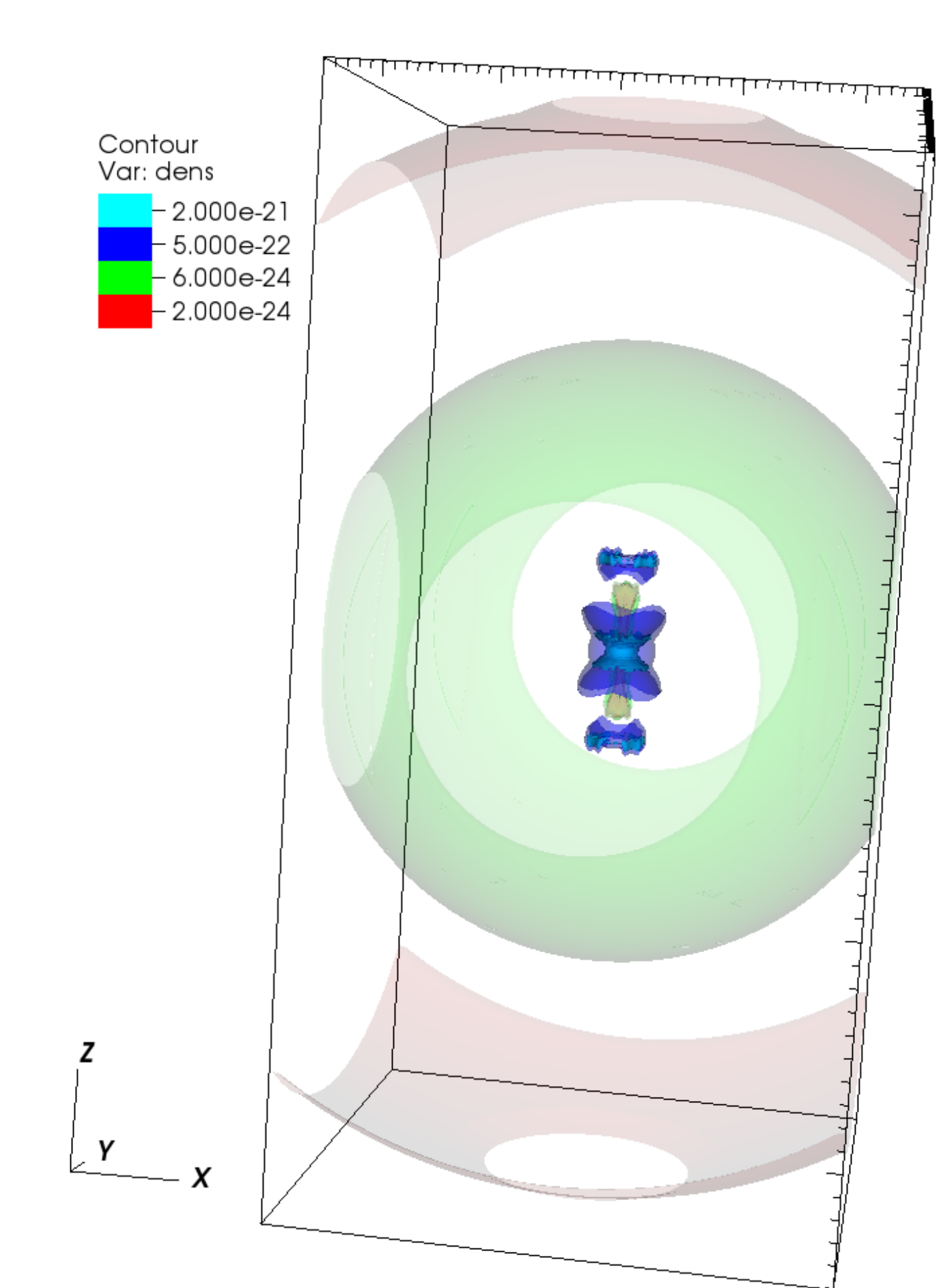}
  \includegraphics[trim= 1.0cm 0.2cm 5.5cm 0.0cm,clip=true,width=0.33\textwidth]{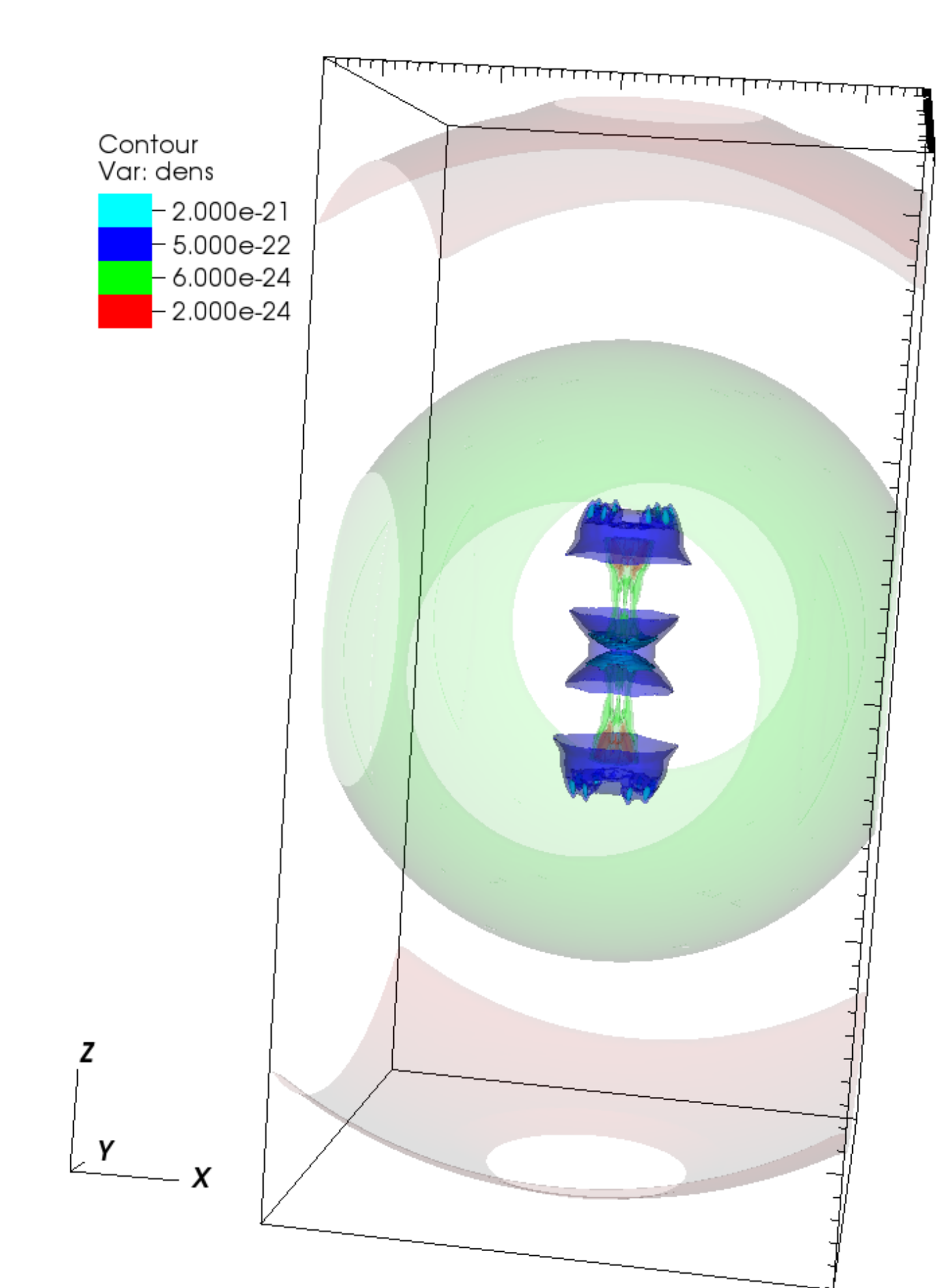}
  \newline
  \vskip 0.3 cm
    \hskip -0.9 cm
 \includegraphics[trim= 1.0cm 0.2cm 5.5cm 0.0cm,clip=true,width=0.33\textwidth]{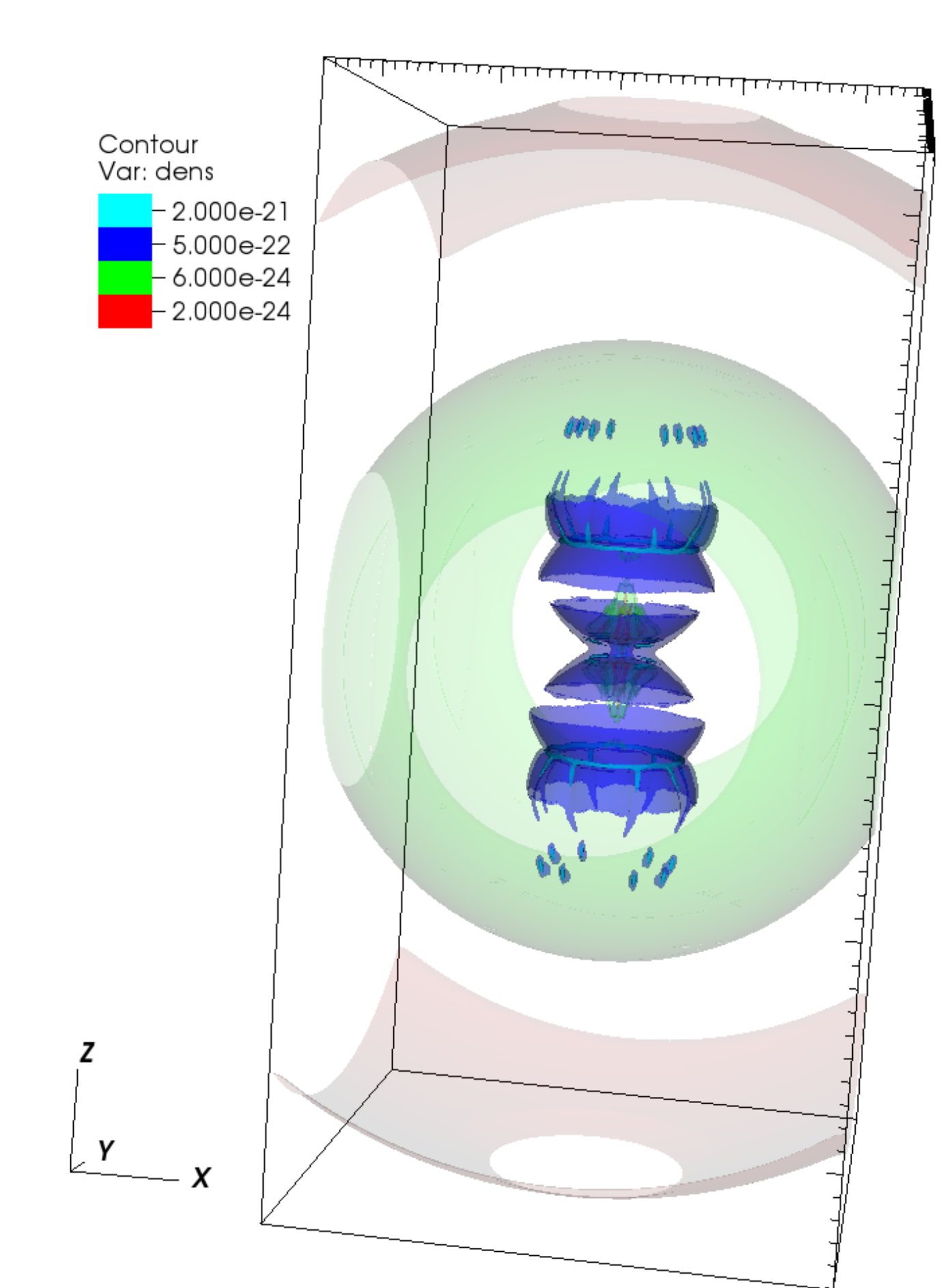}
\includegraphics[trim= 1.0cm 0.2cm 5.5cm 0.0cm,clip=true,width=0.33\textwidth]{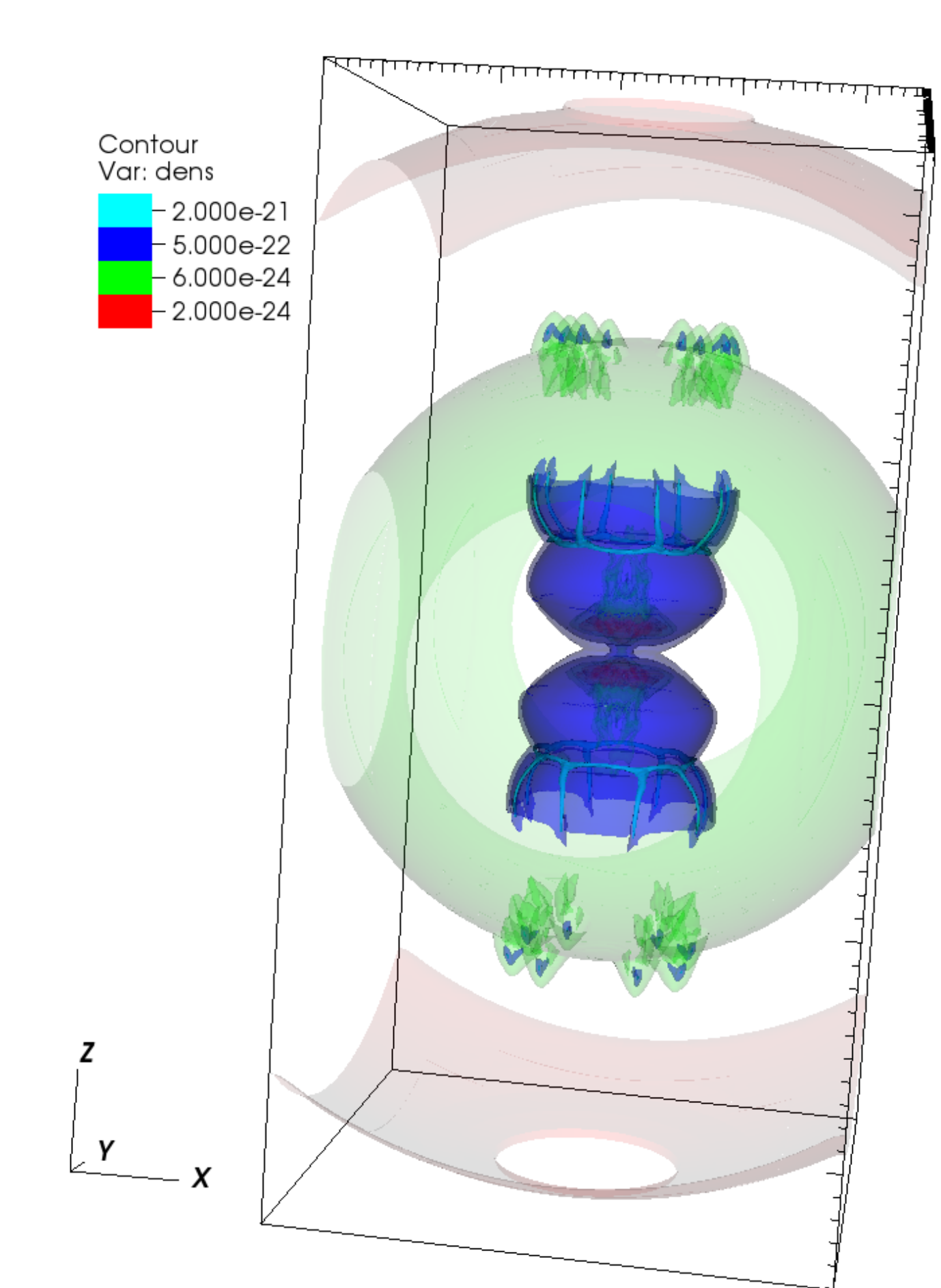}
 \includegraphics[trim= 1.0cm 0.2cm 5.5cm 0.0cm,clip=true,width=0.33\textwidth]{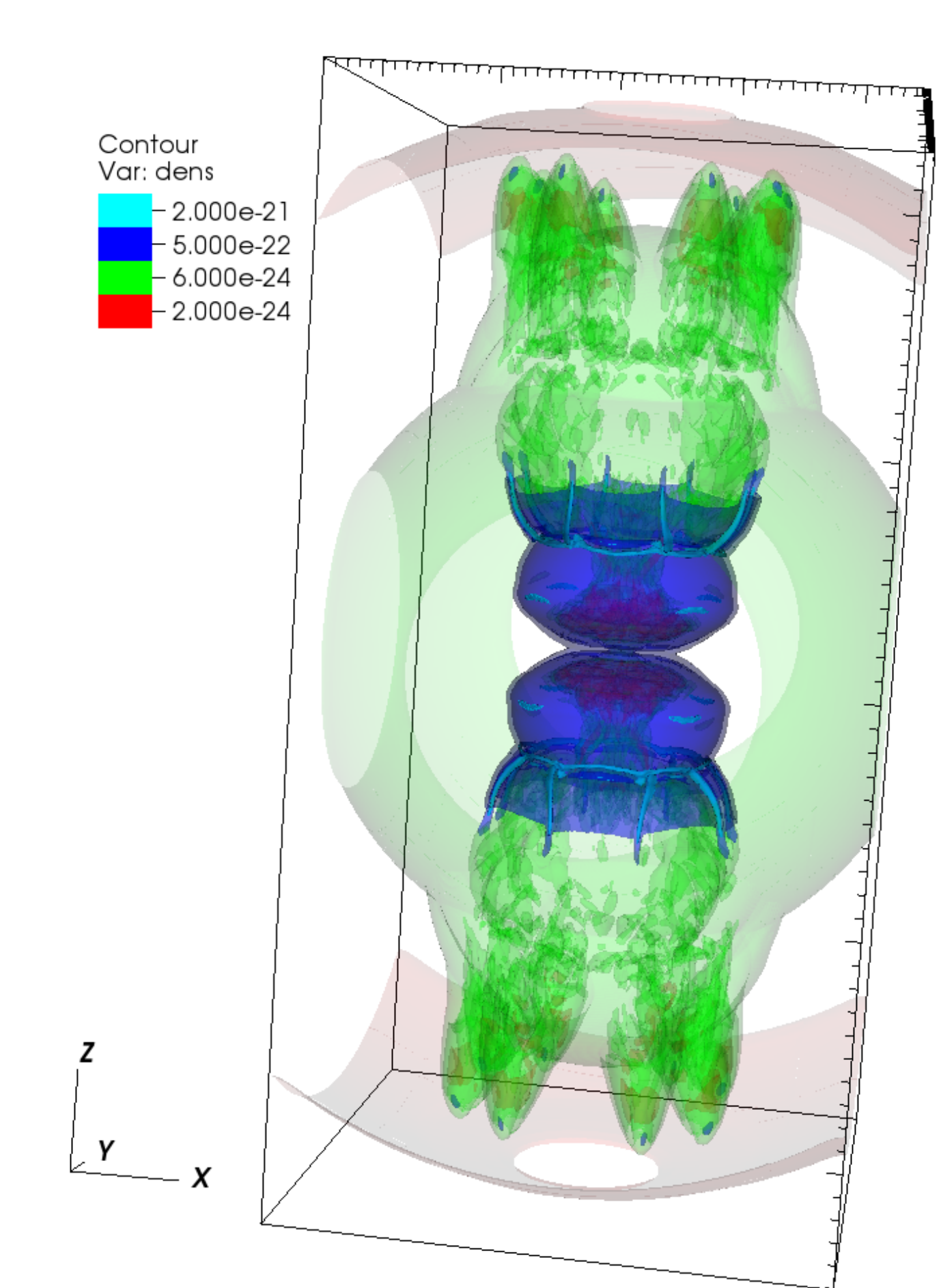}
 \caption{Three dimensional density structure at six times, from upper left to bottom right: $12 \yr$, $44 \yr$, $70 \yr$, $120 \yr$, $177\yr$, and $273\yr$.
 The size of the box is $5\times 10^{17}\cm$ x $5\times 10^{17}\cm$ x $10^{18}\cm$.
 The density scale is given by the colour-bar, where red, green, blue and pale blue represent densities of $2 \times 10^{-24}\g \cm^{-3}$, $6 \times 10^{-24}\g \cm^{-3}$, $5 \times 10^{-22}\g \cm^{-3}$, and $2 \times 10^{-21}\g \cm^{-3}$, respectively.}
  \label{fig:3D}
 \end{center}
 \end{figure}

We start by presenting the 3D structure of the new morphological features that we have obtained here from the particular setting of a slow dense spherical shell embedded in the slow wind, and two jets that interact with the slow spherical ambient gas.   
In Fig. \ref{fig:3D} we present the appearance of the nebula at six times from $t=12 \yr$ to $t=273 \yr$, where $t=0$ is the time when we start to inject the jets. The jets were active for 17 years. The density structure is depicted by four colors. 
  
\begin{figure}
 \begin{center}
  \hskip -0.9 cm
 \includegraphics[trim= 0.0cm 0.0cm 0.0cm 0.0cm,clip=true,width=0.83\textwidth]{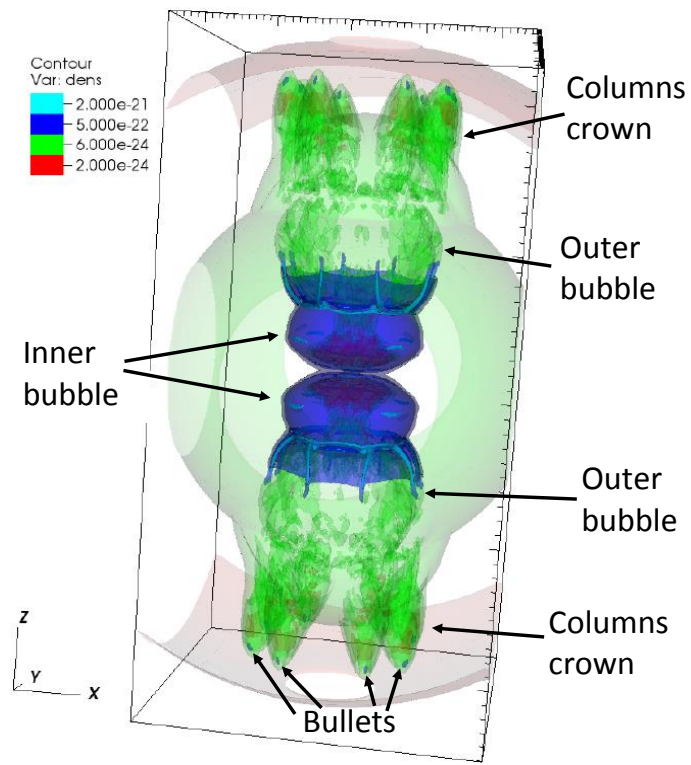}
\vskip -3.6 cm
 \caption{The prominent morphological features that our simulation reveals. }
  \label{fig:Scematic}
 \end{center}
 \end{figure}

Two prominent morphological features develop by the end of the simulation. The first one is a bipolar structure, composed of two opposite lobes, one at each side of the equatorial plane. The structure of each lobe is not a simple prolate or oblate shape, but rather we see two bubbles on each side, as we mark on Fig. \ref{fig:Scematic}. A bubble with a denser surface that is seen in blue touches the center on the close side to the center, and is connected to a second bubble on the far side. In the panel at $t=273 \yr$ of Fig. \ref{fig:3D} the outer bubble is seen in blue in its close part, and in green in its far side from the center. Dense thin filaments are seen in blue on the surface of the outer bubble. They extend from the boundary between the two bubbles and up to the middle of the outer bubble.  

The second prominent feature, which is the focus of this paper, is the structure of straight eight wide filaments, or columns, that are seen in green at $t=273 \yr$ in Fig. \ref{fig:3D}, extending more or less in the $z$ direction from the outer bubble out. We term this structure columns crown. As we show below, they develop from Rayleigh-Taylor instability modes. We do note that here we form 8 thick columns. The numerical grid determines the number and width of the columns. Later we raise the possibility that in reality many more columns and thiner ones are formed.      
  
\begin{figure}
 \begin{center}
\hskip -0.3 cm
 \includegraphics[trim= 1.0cm 0.2cm 7.5cm 0.0cm,clip=true,width=0.33\textwidth]{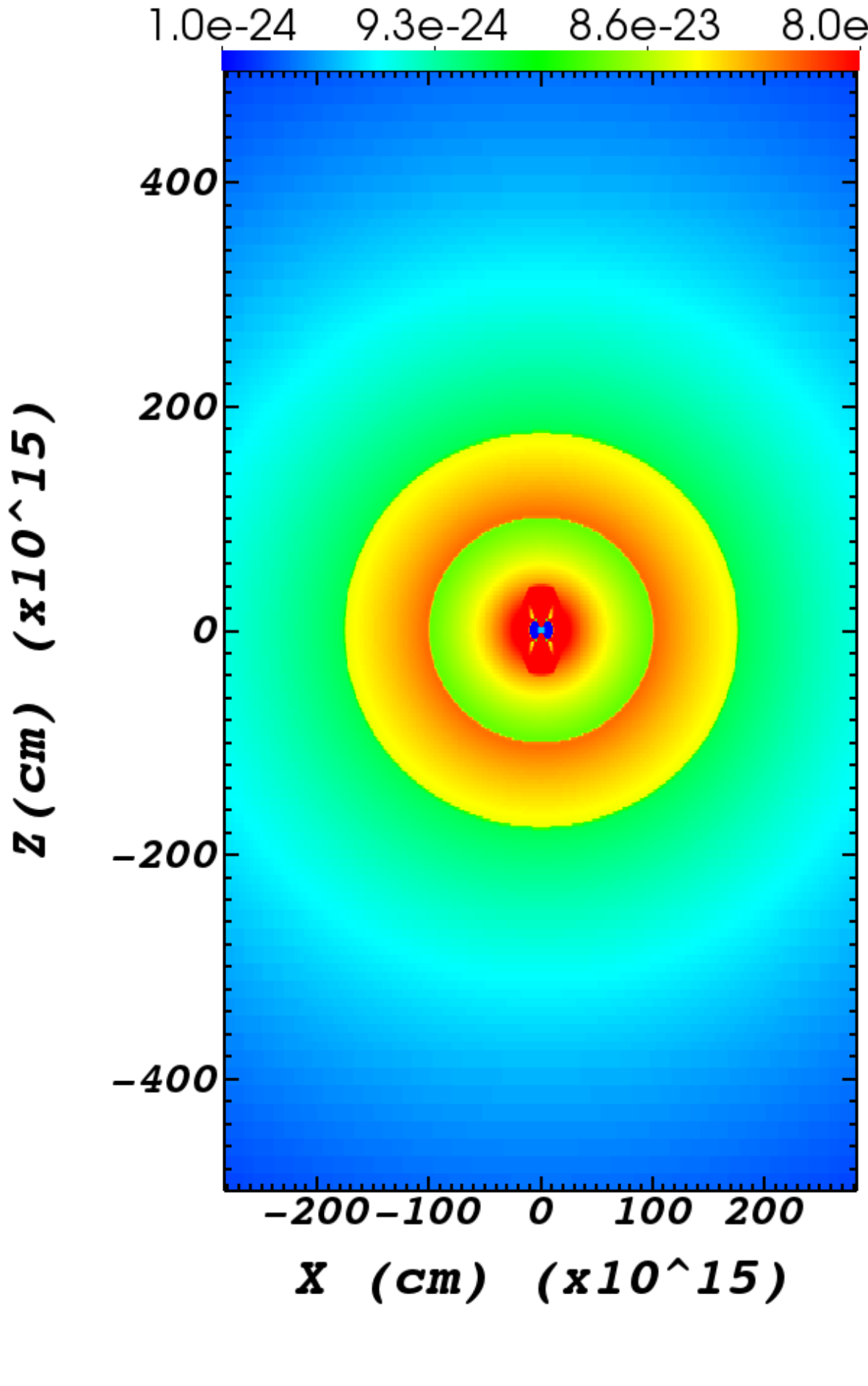}
 \includegraphics[trim= 1.0cm 0.2cm 7.5cm 0.0cm,clip=true,width=0.33\textwidth]{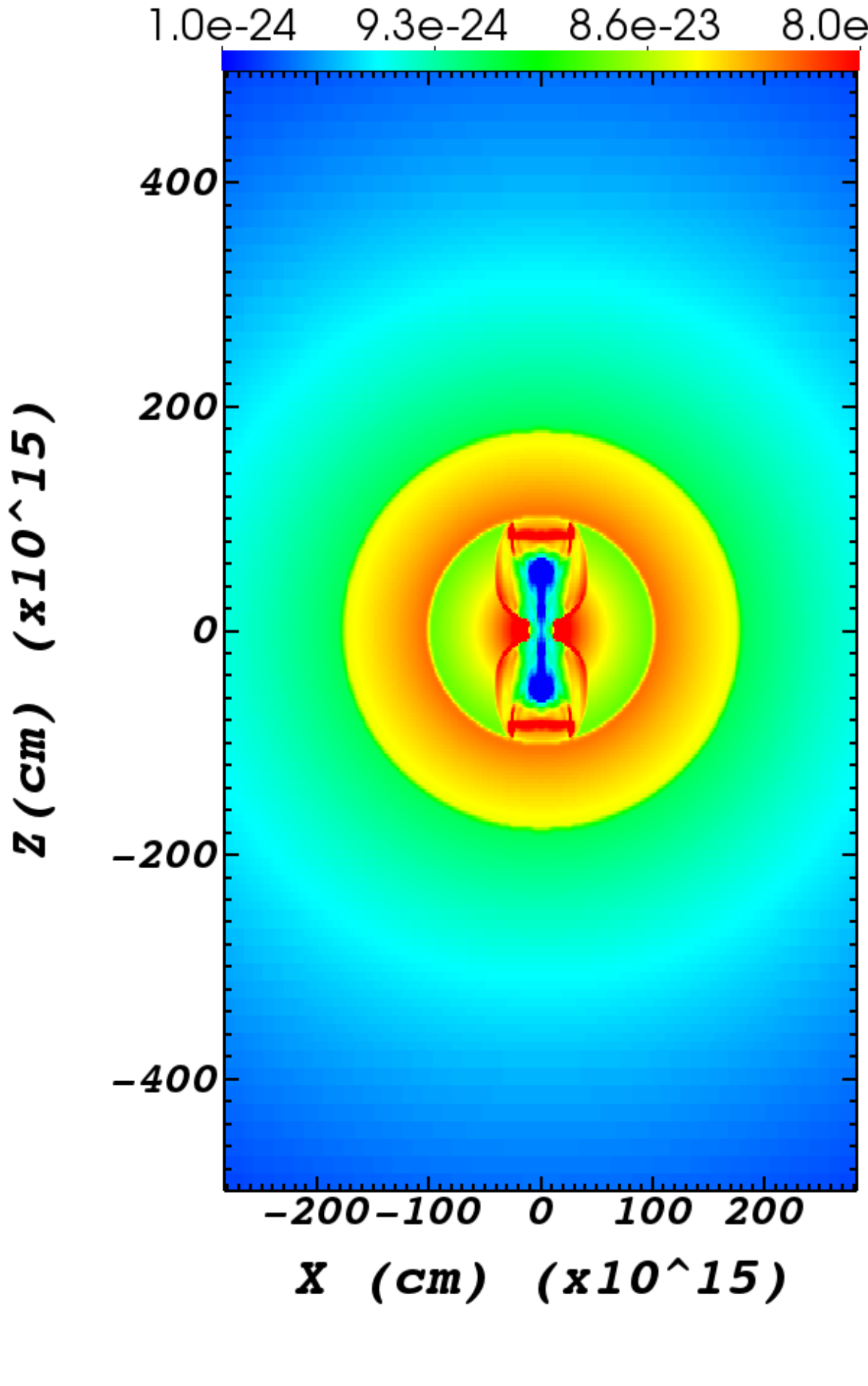}
 \includegraphics[trim= 1.0cm 0.2cm 7.5cm 0.0cm,clip=true,width=0.33\textwidth]{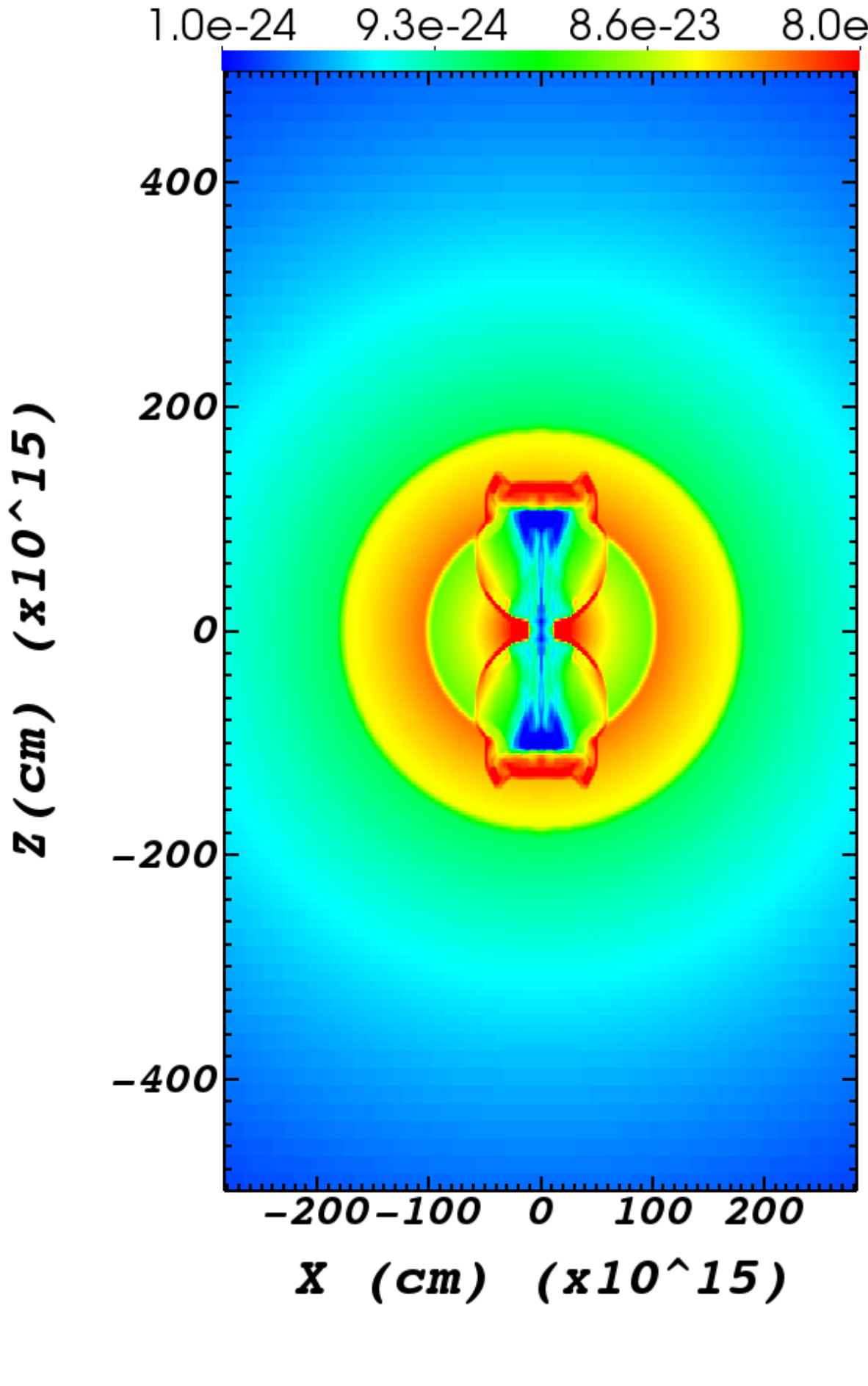}
\end{center}
 \begin{center}
 \hskip -0.3 cm
 \includegraphics[trim= 1.0cm 0.2cm 7.5cm 0.0cm,clip=true,width=0.33\textwidth]{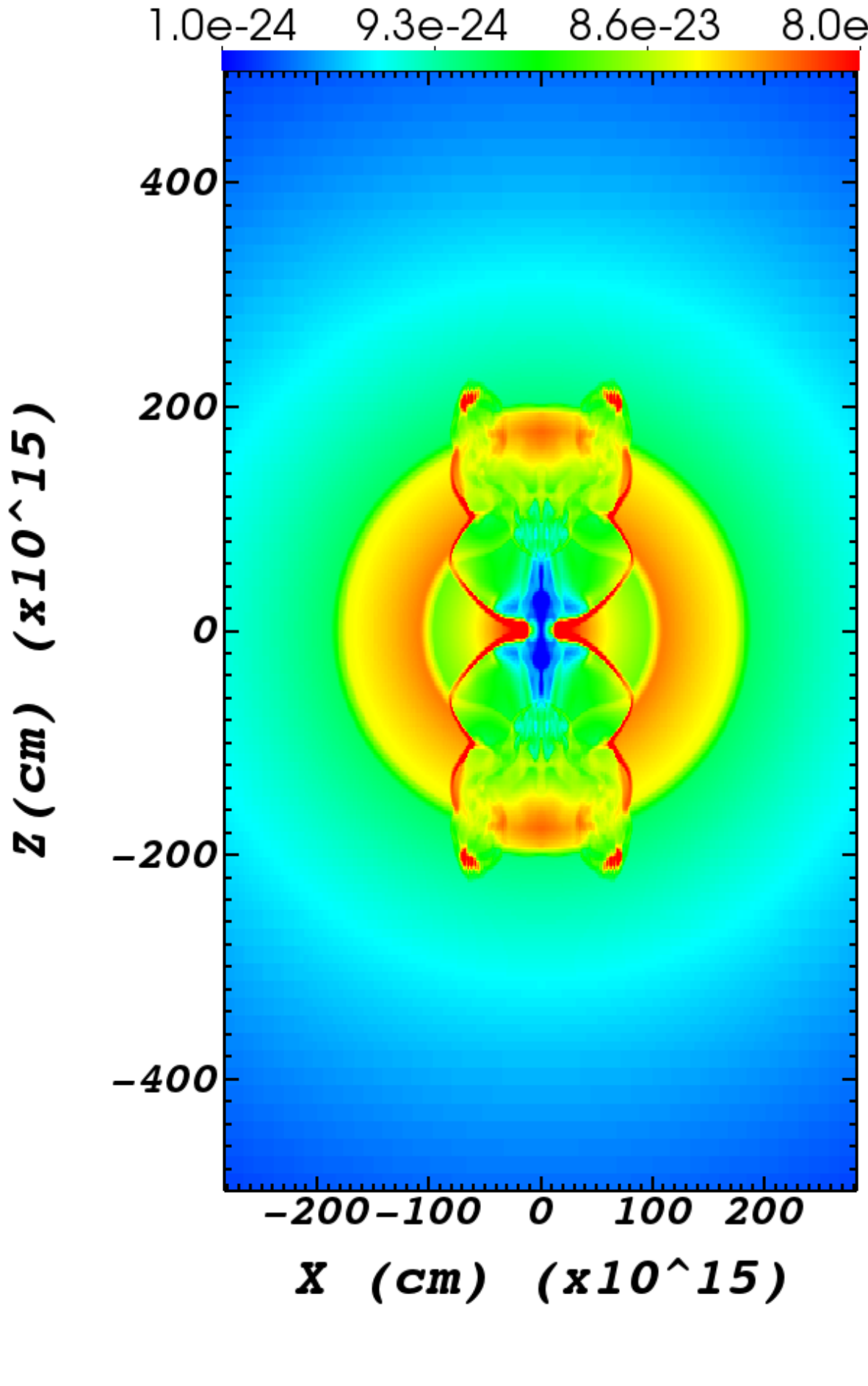}
 \includegraphics[trim= 1.0cm 0.2cm 7.5cm 0.0cm,clip=true,width=0.33\textwidth]{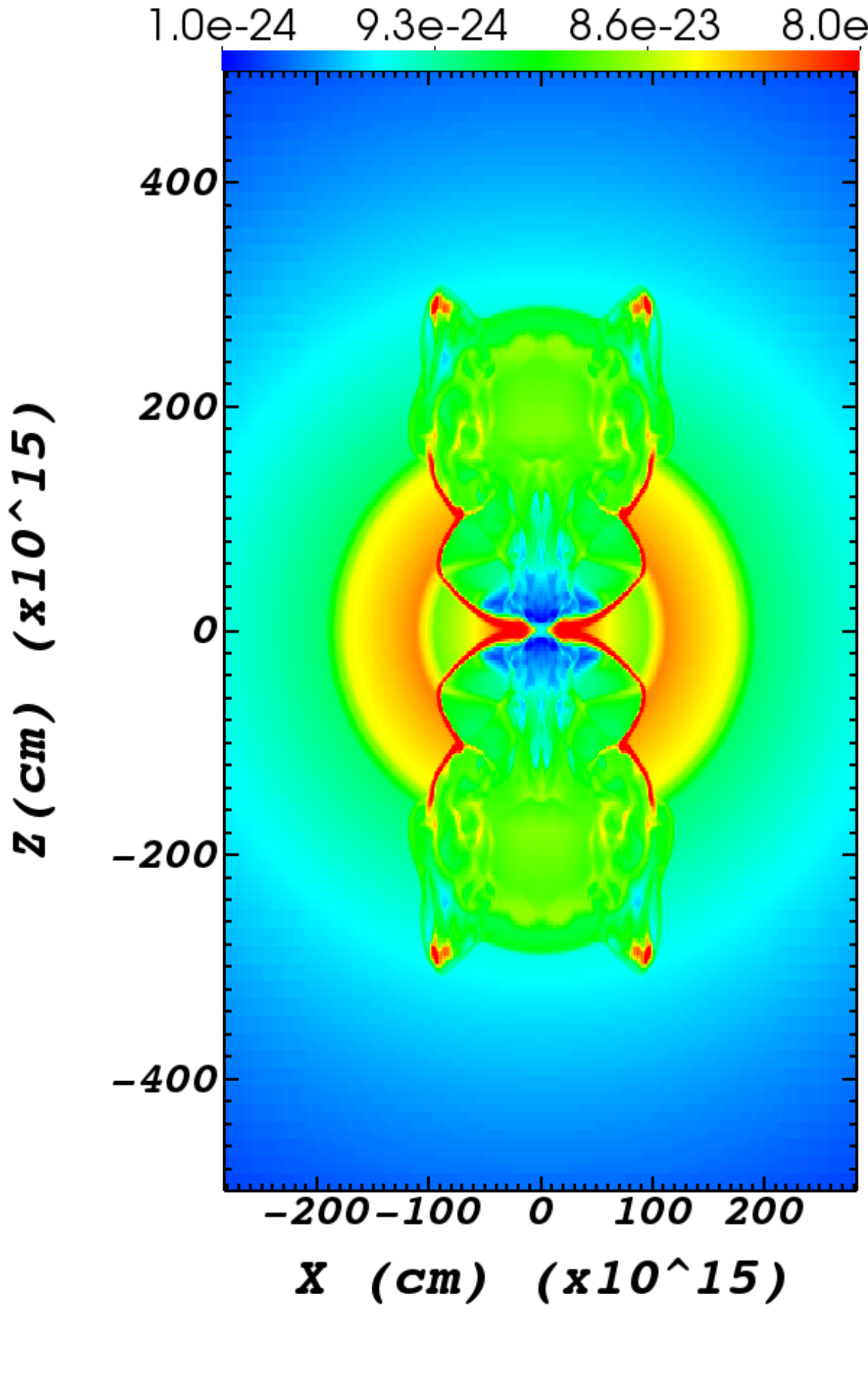}
 \includegraphics[trim= 1.0cm 0.2cm 7.5cm 0.0cm,clip=true,width=0.33\textwidth]{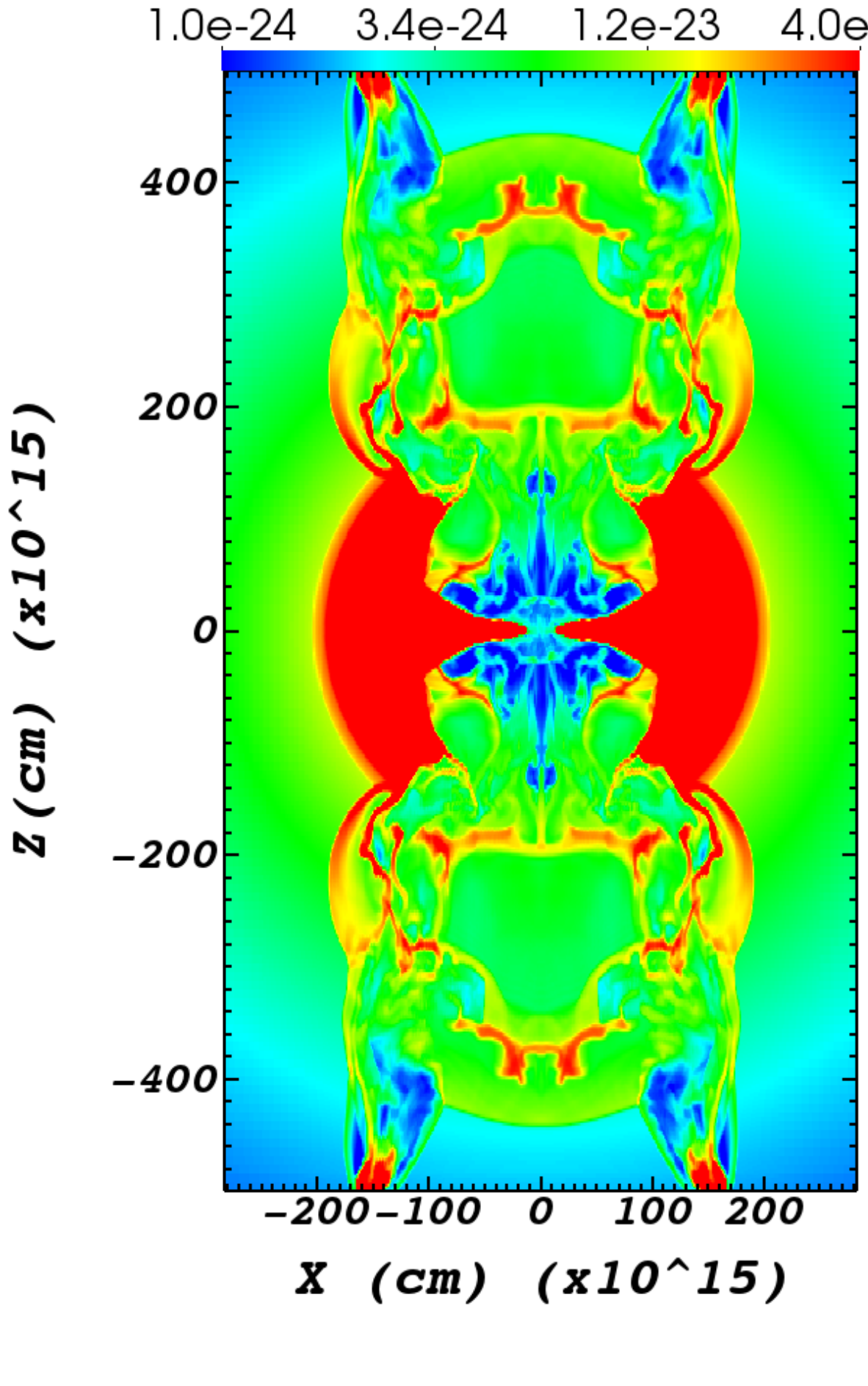}
 \vskip -0.6 cm
 \caption{The density maps in a meridional plane that is rotated at $60^\circ$ from the plane $x=0$. The horizontal axis is the distance along the line $y=x \tan 60^\circ$. We present the density at six times: $12 \yr$, $44 \yr$, $70 \yr$, $120 \yr$, $177$, and $317\yr$. The density scale is given by the colour-bar in units of $\g \cm^{-3}$. In the first 5 panels the red color represents to a density of $8 \times 10^{-22}\g \cm^{-3}$, while in the last panel the red color represents a density of $4 \times 10^{-23}\g \cm^{-3}$. } 
  \label{fig:dens_slice}
 \end{center}
 \end{figure}

\begin{figure}
 \begin{center}
 \hskip -0.3 cm
\includegraphics[trim= 1.0cm 0.2cm 7.5cm 0.0cm,clip=true,width=0.33\textwidth]{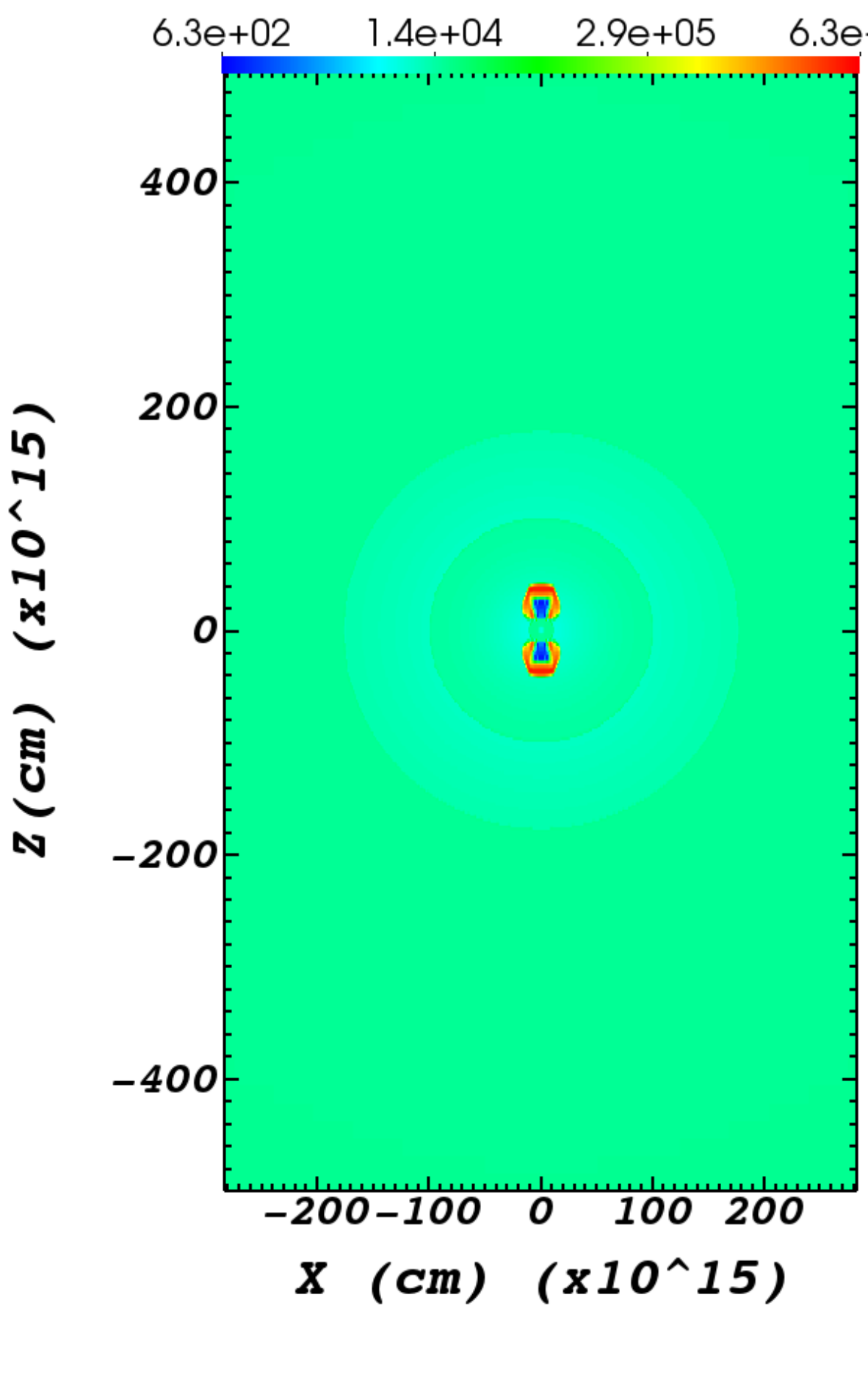}
 \includegraphics[trim= 1.0cm 0.2cm 7.5cm 0.0cm,clip=true,width=0.33\textwidth]{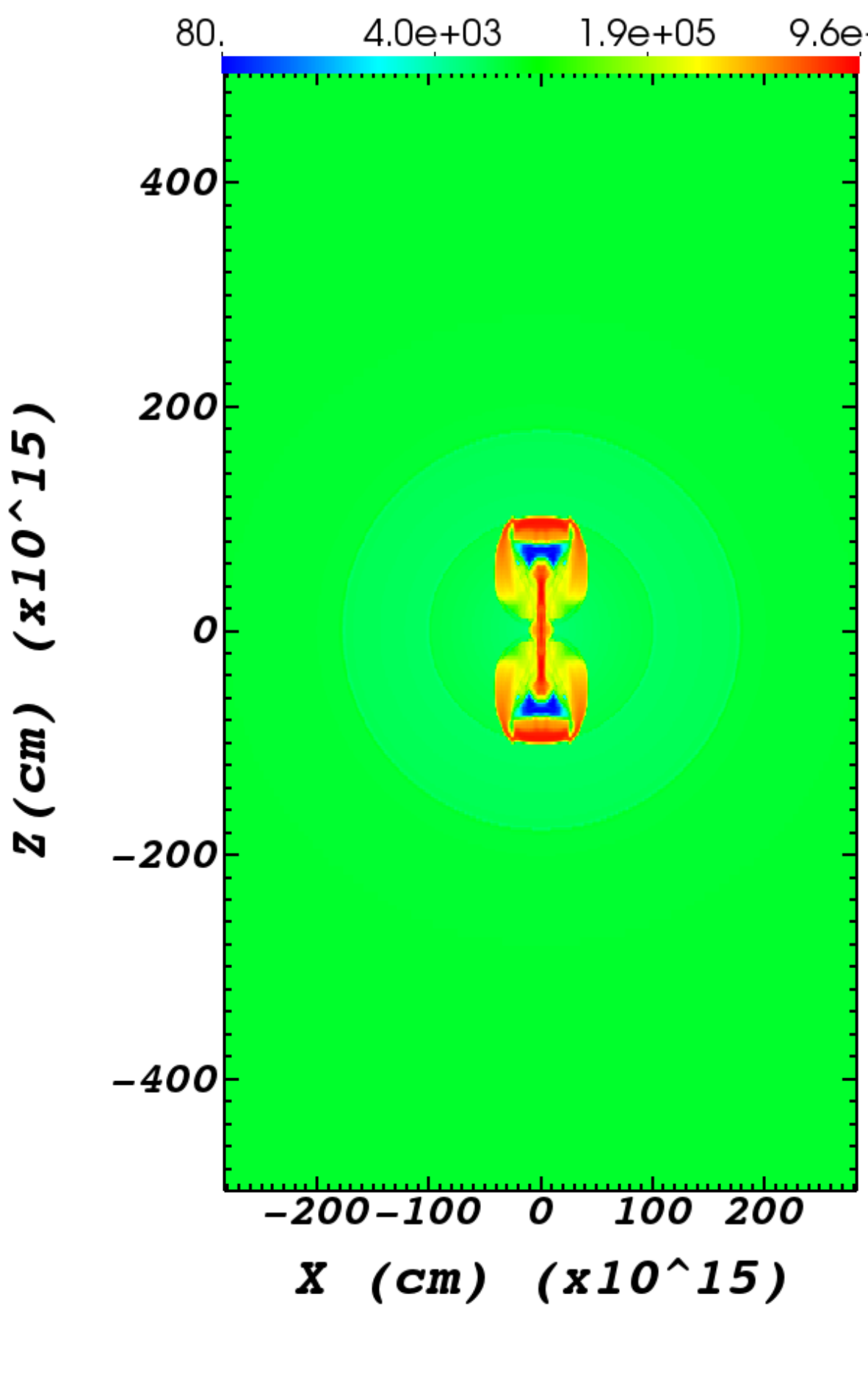}
\includegraphics[trim= 1.0cm 0.2cm 7.5cm 0.0cm,clip=true,width=0.33\textwidth]{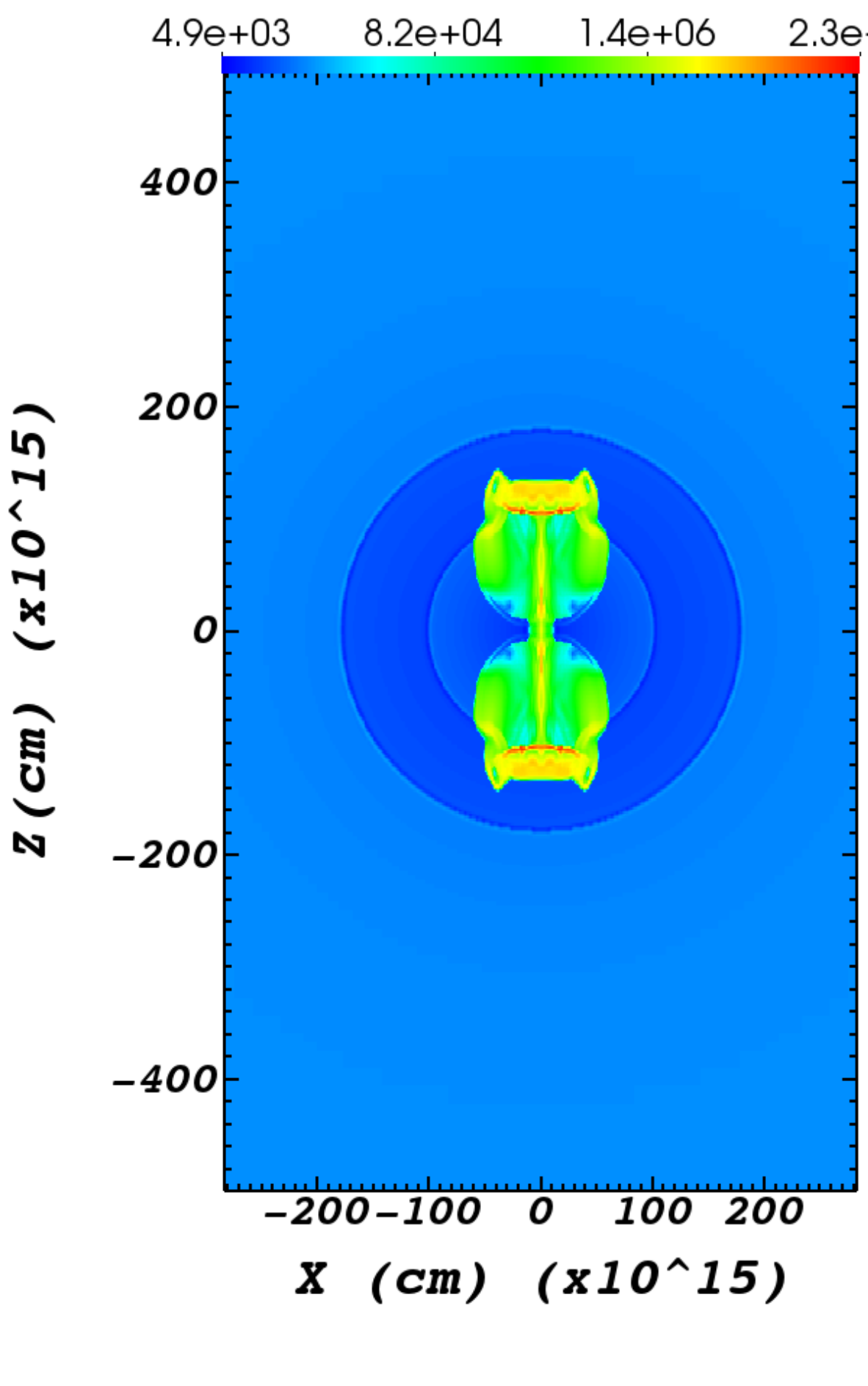}
 \end{center}
 \begin{center}
\hskip -0.3 cm
 \includegraphics[trim= 1.0cm 0.2cm 7.5cm 0.0cm,clip=true,width=0.33\textwidth]{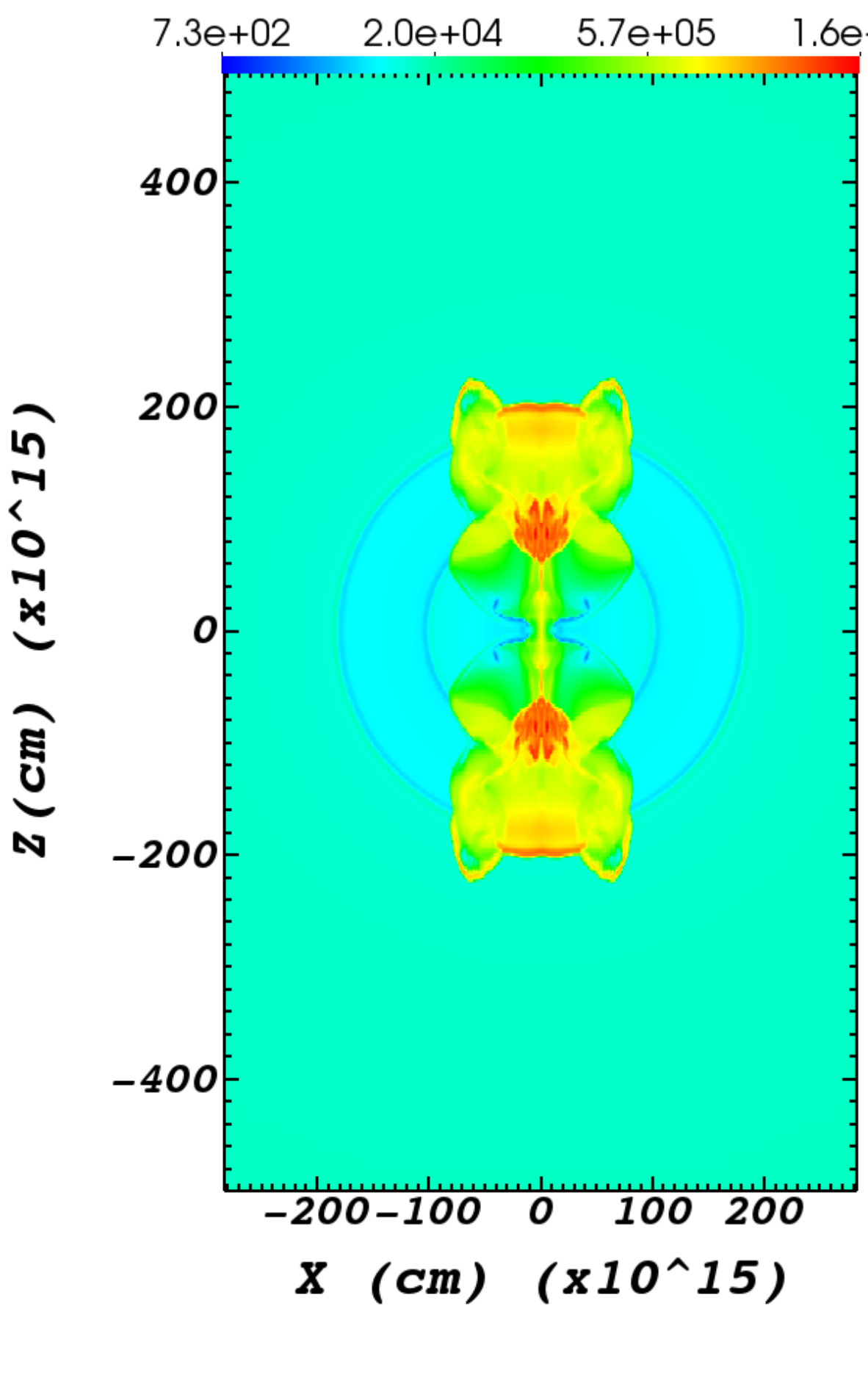}
 \includegraphics[trim= 1.0cm 0.2cm 7.5cm 0.0cm,clip=true,width=0.33\textwidth]{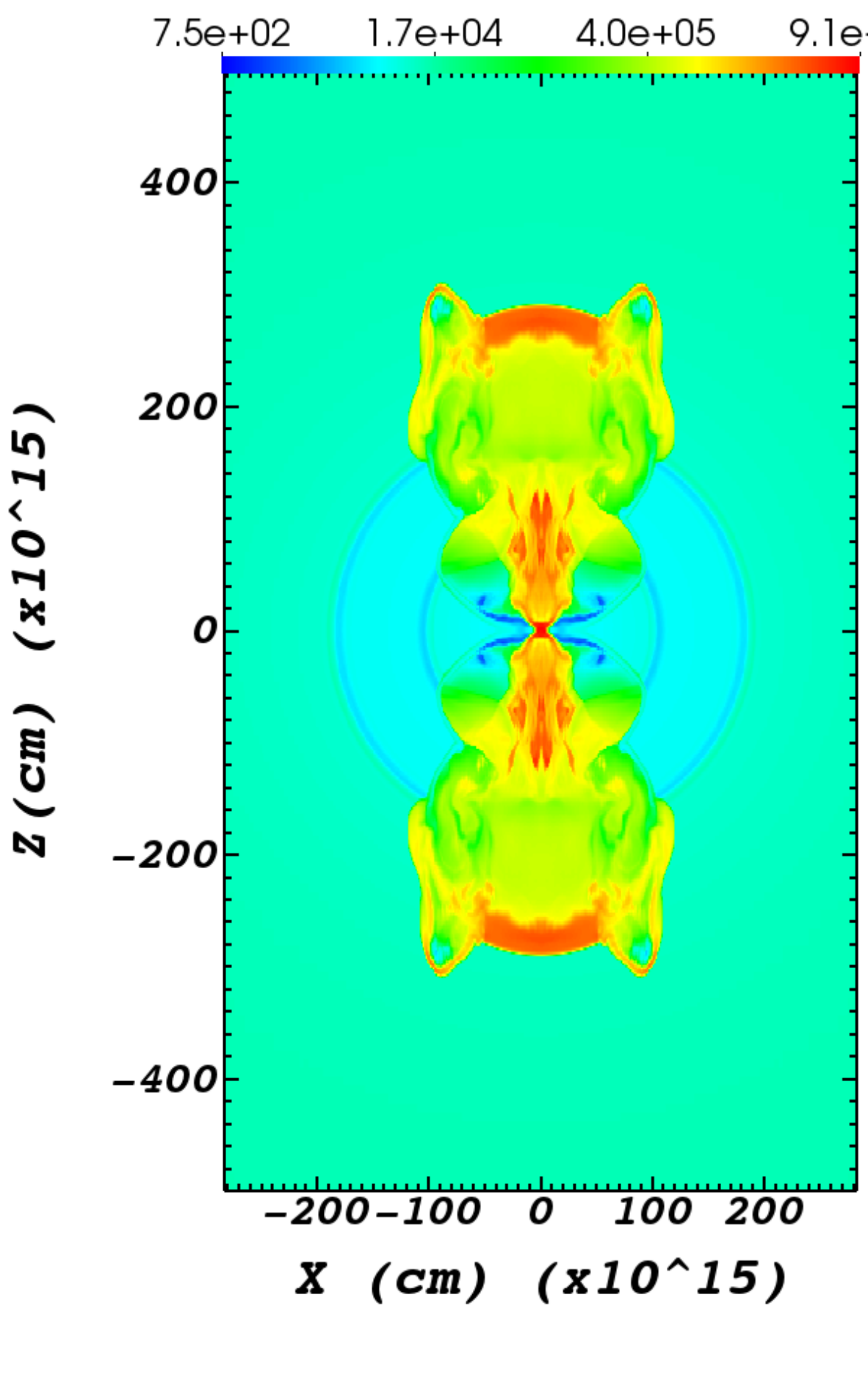}
\includegraphics[trim= 1.0cm 0.2cm 7.5cm 0.0cm,clip=true,width=0.33\textwidth]{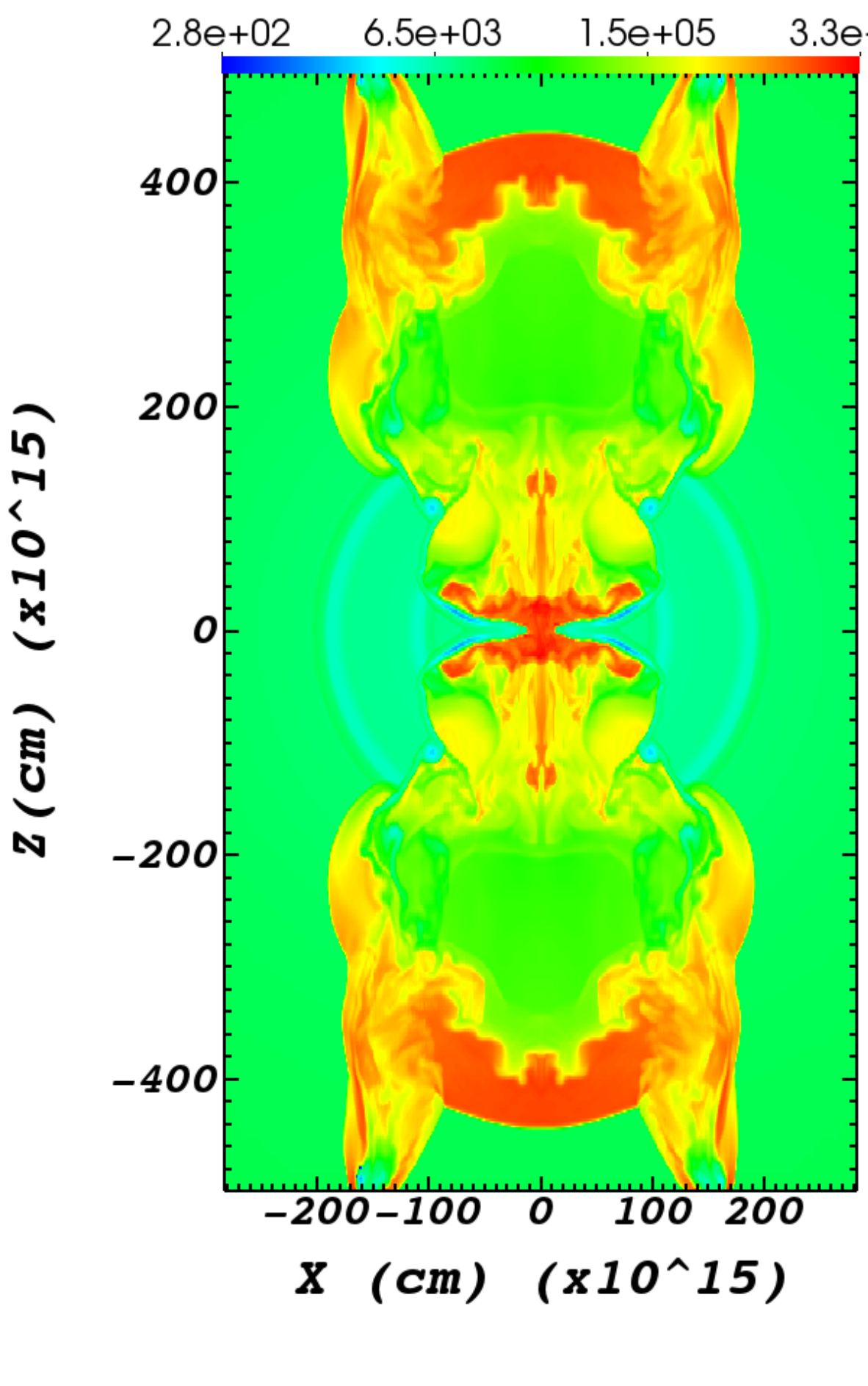}
\vskip -0.6 cm
 \caption{The temperature in the same plane and at the same times as in Fig. \ref{fig:dens_slice}. The Temperature scale is given by the colour-bar in units of $\K$ (the red color is in units of $10^6$; this was trimmed from the color bar). }
  \label{fig:temp_slice}
 \end{center}
 \end{figure}

\begin{figure}
 \begin{center}
 \hskip -0.3 cm
\includegraphics[trim= 1.0cm 0.2cm 7.5cm 0.0cm,clip=true,width=0.33\textwidth]{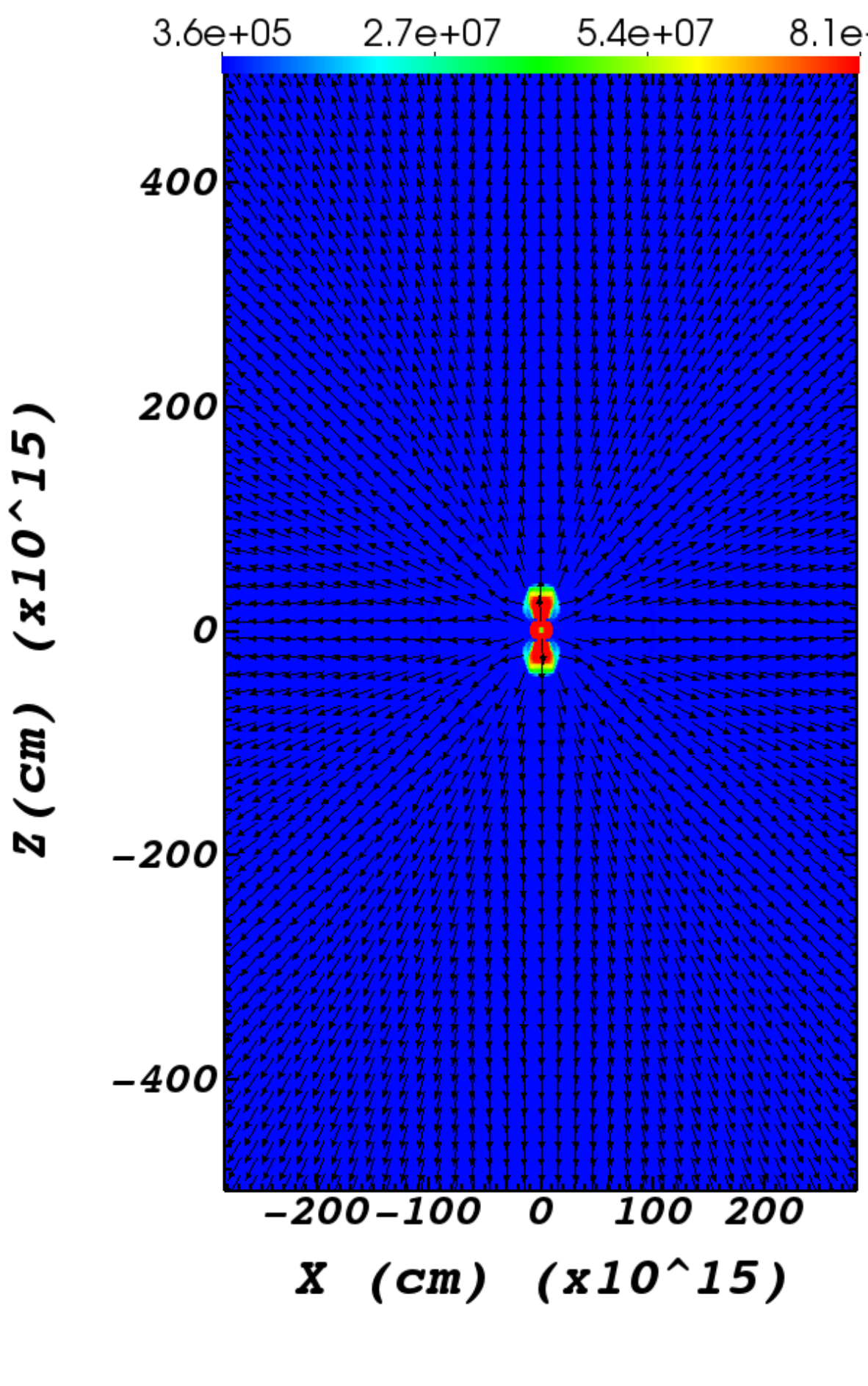}
\includegraphics[trim= 1.0cm 0.2cm 7.5cm 0.0cm,clip=true,width=0.33\textwidth]{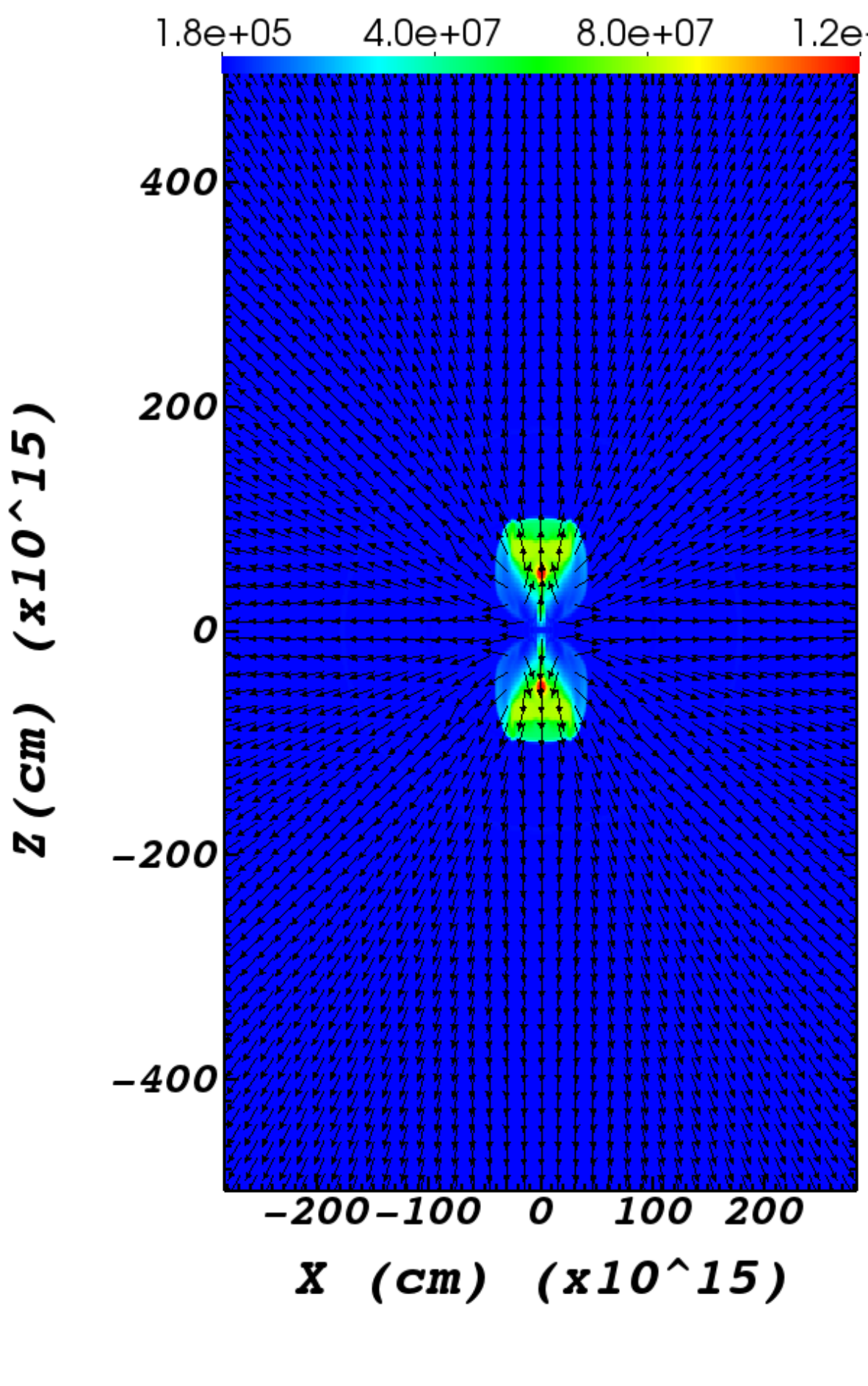}
\includegraphics[trim= 1.0cm 0.2cm 7.5cm 0.0cm,clip=true,width=0.33\textwidth]{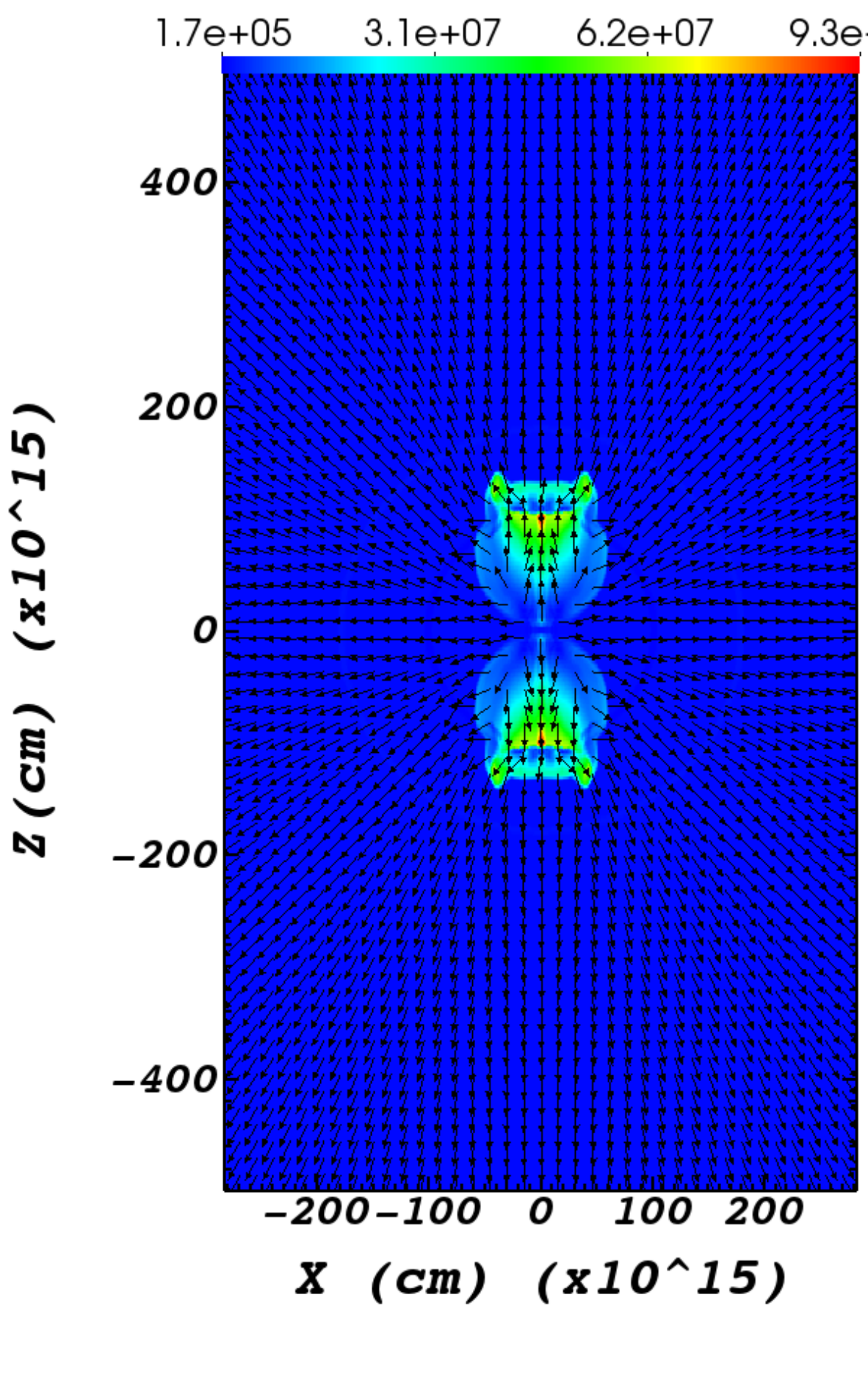}
 \end{center}
 \begin{center}
\hskip -0.3 cm
 \includegraphics[trim= 1.0cm 0.2cm 7.5cm 0.0cm,clip=true,width=0.33\textwidth]{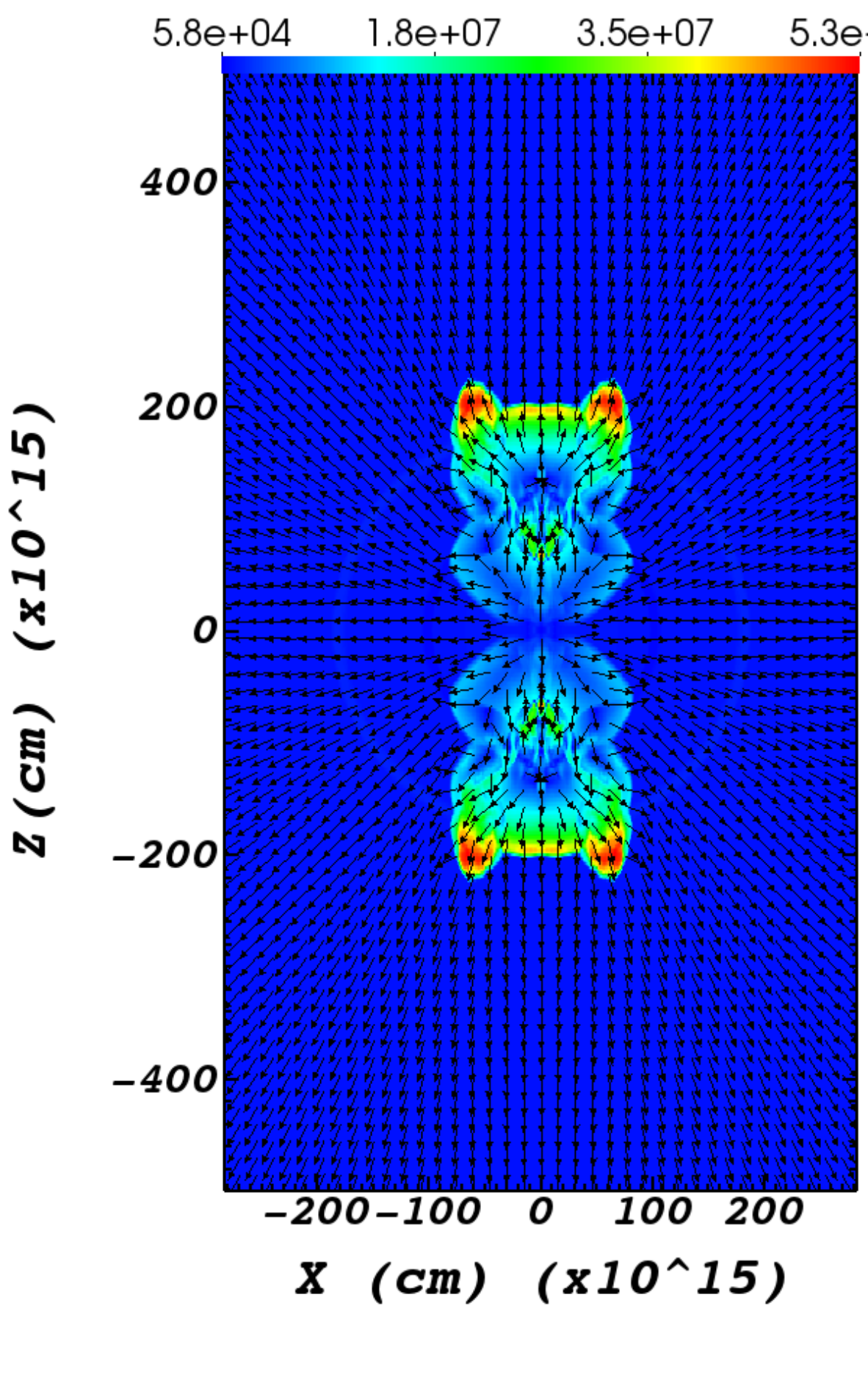}
\includegraphics[trim= 1.0cm 0.2cm 7.5cm 0.0cm,clip=true,width=0.33\textwidth]{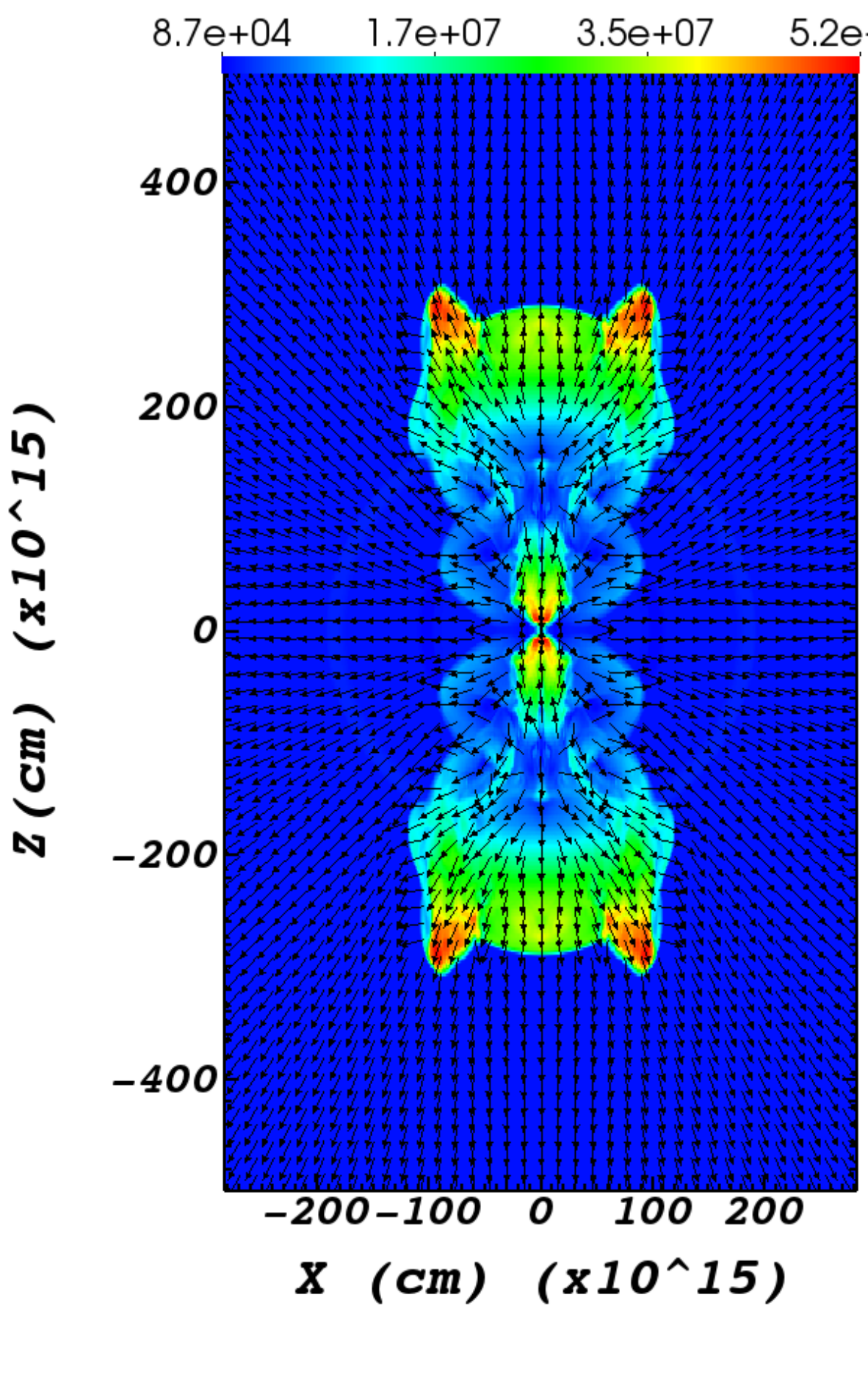}
\includegraphics[trim= 1.0cm 0.2cm 7.5cm 0.0cm,clip=true,width=0.33\textwidth]{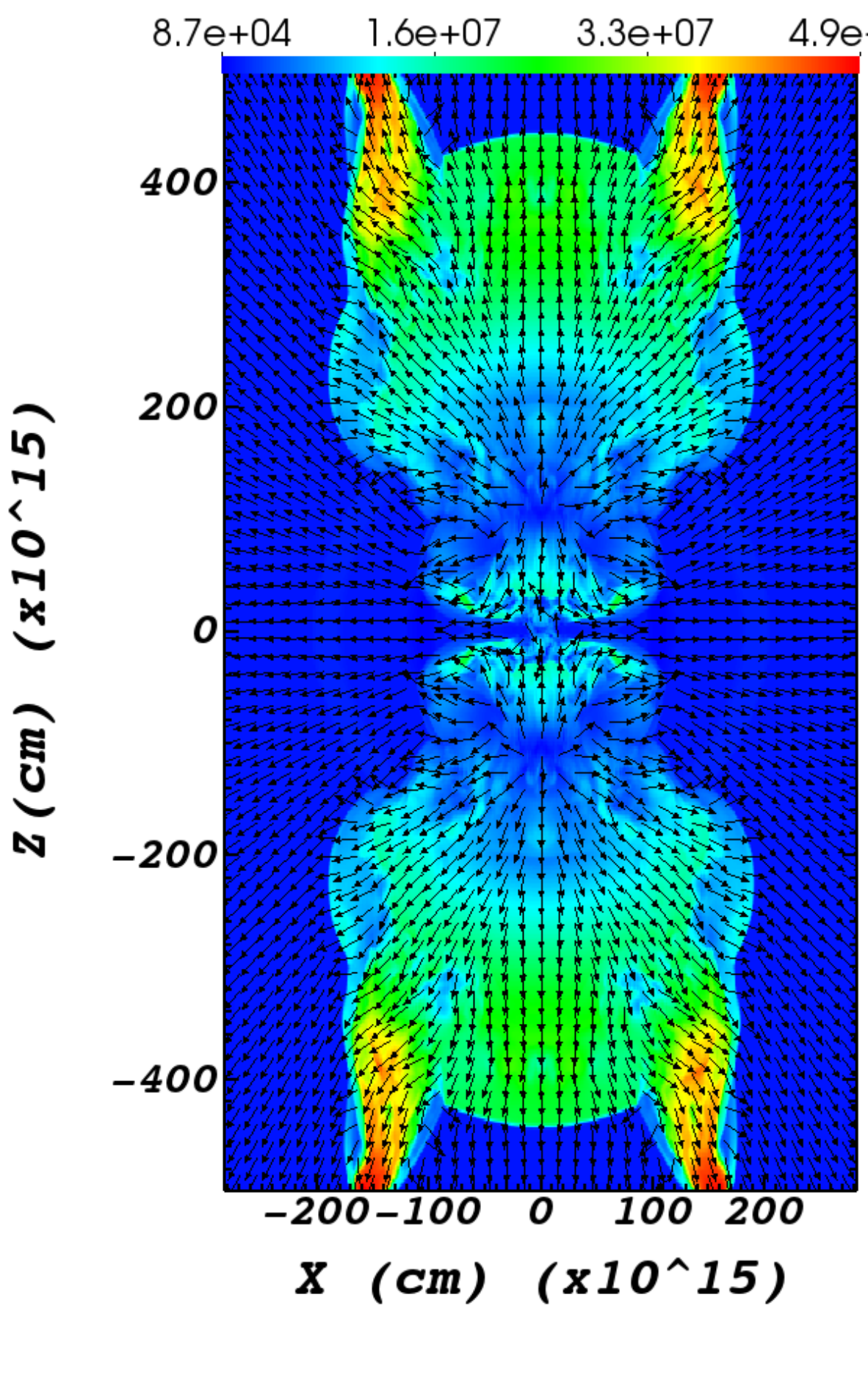}
 \vskip -0.6 cm
 \caption{The velocity in the same plane and same times as in Fig. \ref{fig:dens_slice}. Arrows present the direction of velocity and colors present the magnitude of the velocity according to the colour-bar and in units of $\cm \s^{-1}$ (the red color is in units of $10^7$, beside the second panel where it is in units of $10^8$). }
 \label{fig:vel_arrows}
 \end{center}
 \end{figure}

 In Fig. \ref{fig:dens_slice} -  \ref{fig:vel_arrows} we present the density, temperature, and velocity maps, respectively, in a meridional plane at 6 times. This meridional plane has an angle of $60^\circ$ to the plane $x=0$. We choose this meridional plane to include two of the columns from the column crown, one column at each side of the symmetry axis $z$. The Horizontal axis in the figures where we present this meridional plane is the distance along the line $y=x \tan 60^\circ$.      We can discern the following interaction phases. 

\textit{1. Jet injection and early interaction.} 
We inject the jets only in the first 17 years. The upper left panel in each of the figures, at $t=12 \yr$, depicts this phase. At this phase we can still see the three parts of the spherical structure that the giant star formed according to our initial setting (section \ref{sec:numerical}), all expanding out at a velocity of $10 \km \s^{-1}$. A slow wind fills the volume from the center to $r=10^{17} \cm$, a denser shell from $r=10^{17} \cm$ to $r=1.75 \times 10^{17} \cm$ that was formed from an eight times higher mass loss rate episode, and a slow wind that extends from the dense shell to the edge of the numerical grid. The jets just start to inflate a bipolar structure.   

\textit{2. Early inflation of bipolar nebula.}  In the first 45 years, that are depicted by the first two panels of Fig. \ref{fig:dens_slice} -  \ref{fig:vel_arrows}, the jets inflate a bipolar structure in the slow wind that was blown by the central star after the formation of the dense shell. 
The jets' gas passes through shock waves and form hot zones with a temperature of up to $\simeq 10^7 \K$. The density structure at $t=44 \yr$ (second panel) reveals that the dense front of the inflated bipolar structure is a plane (red horizontal bar on each side of the equatorial plane), but that it has two protrusions, one at its edge of this bar. These will later develop into columns and be part of the columns crown.   

\textit{3. Interaction with the dense shell.} The two panels that come next, at $t=70 \yr$ and $t=120 \yr$,show the outcome of the interaction with the dense shell. The panel at $t=120 \yr$ shows that the interaction of the expanding bipolar structure with the dense shell forms in each side of the equatorial plane an outer bubble. The lobe that was inflated earlier forms now a bubble closer to the center, while the interaction with the dense shell forms an outer bubble that is connected to the one closer to the center. The boundary between the two bubbles coincides with the inner radius of the dense shell at $10^{17} \cm$. The outer bubble on each side of the new bipolar structure opens up on its far (from the center) side, and protrusions extend further out. These protrusions, that are Rayleigh-Taylor instability tongues, form the columns crown.

Only a small fraction of the mass in the dense shell lies along the path of the jets and the bipolar structure they form when interacting with the inner slow wind. We turned off the jets before this bipolar structure hits the dense shell, so it is the bipolar structure that fractures the dense shell. The interaction forms two bubbles, the outer one that forms the columns crown comes from the dense shell.  
At the end of our simulation the outflow removed two caps, one at each side of the equatorial plane, from the dense shell. The half opening angle of each of the removed caps is about $40 ^\circ$, such that the total mass in the two caps is about 23 per cent of the dense shell mass, or about $5 \times 10^{-4} M_\odot$. This is about 15 times the mass in the two jets. 
We note though, that the columns crown is formed at an earlier time, and the fraction of the mass of the dense shell that leads to the formation of the dense crown is only about 10 per cent of the mass of the dense shell (a half opening angle of $25^\circ$), or about $2 \times 10^{-4} M_\odot$.

\textit{4. Late evolution.} 
At later times the bipolar structure expands into the outer slow wind. Now the outcomes of the instabilities become prominent, in particular the columns crown and the filaments and blobs inside the outer bubble of the bipolar structure. 
 In the meridional plane that we present in Fig. \ref{fig:dens_slice}, two columns are seen as two red spots in the very upper boundary of the grid, and two red spots with their columns are seen in the very lower boundary of the grid at $t=317 \yr$ (lower right panel). The red spots are dense bullets that we mark also on Fig. \ref{fig:Scematic}. The higher density columns (they are in green) that extent from the outer bubble to the red spots in Fig. \ref{fig:dens_slice} are the columns that we present in Fig. \ref{fig:3D}. The columns are clearly seen also in the lower right panel of Figs. \ref{fig:temp_slice} and \ref{fig:vel_arrows}, where they have the appearance of `ears'. These `ear' structures are formed by the bullets (shrapnel) that move through the ambient gas. The high temperature that forms the `ear' structure behind each bullet results from a shock wave, as the bullets move at a velocity of $\simeq 500 \km \s^{-1}$ that is highly supersonic. 
The bullets (shrapnel) themselves expand at an angle of about $25^\circ$ to the symmetry axis with a speed of about $500 \km \s^{-1}$, and the mass in each bullet is about  several$~\times 10^{-7} M_\odot$.
 
The red zones inside the outer bubble in Fig. \ref{fig:dens_slice} were formed by instabilities, as we elaborate on in the next section.

Finally, we note that because we do not inject a strong wind or jets from the center anymore, the high pressure in the inner bubble pushes material back to the center. This back-flow turns to an outflow in the equatorial plane as seen in the last panel of Fig. \ref{fig:vel_arrows}. In an earlier paper \citep{AkashiSoker2008b} we studied a different kind of back-flow that the jets and the high pressure bubbles that they inflate form. In general , the possibility of a gas that flows back to the center is an interesting subject for future studies. 

\subsection{Instabilities and the columns crown}
\label{subsec:instabilities}
 
\begin{figure}
\includegraphics[trim= 1.0cm 0.2cm 7.5cm 0.2cm,clip=true,width=0.33\textwidth]{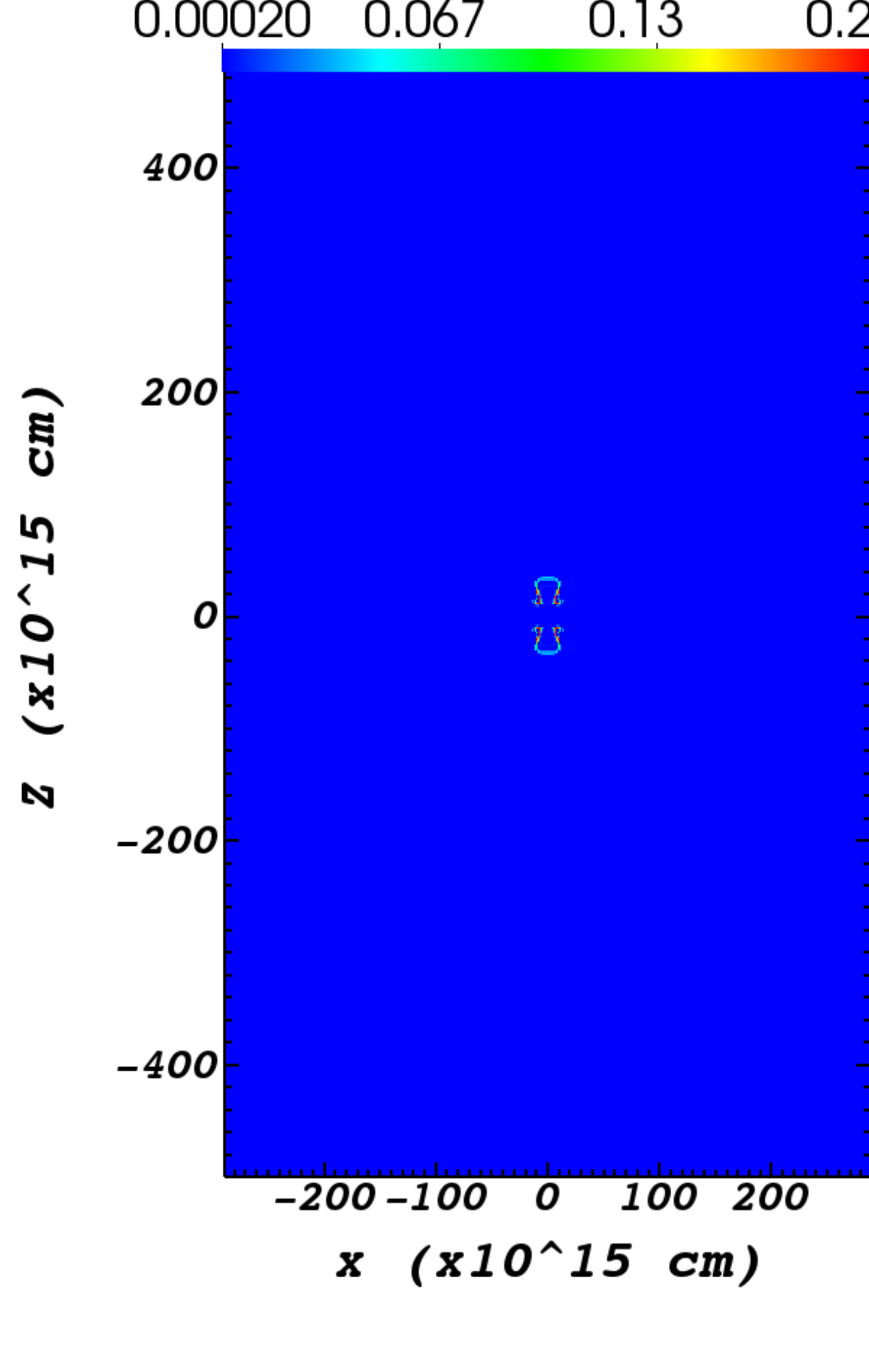}
\includegraphics[trim= 1.0cm 0.2cm 7.5cm 0.2cm,clip=true,width=0.33\textwidth]{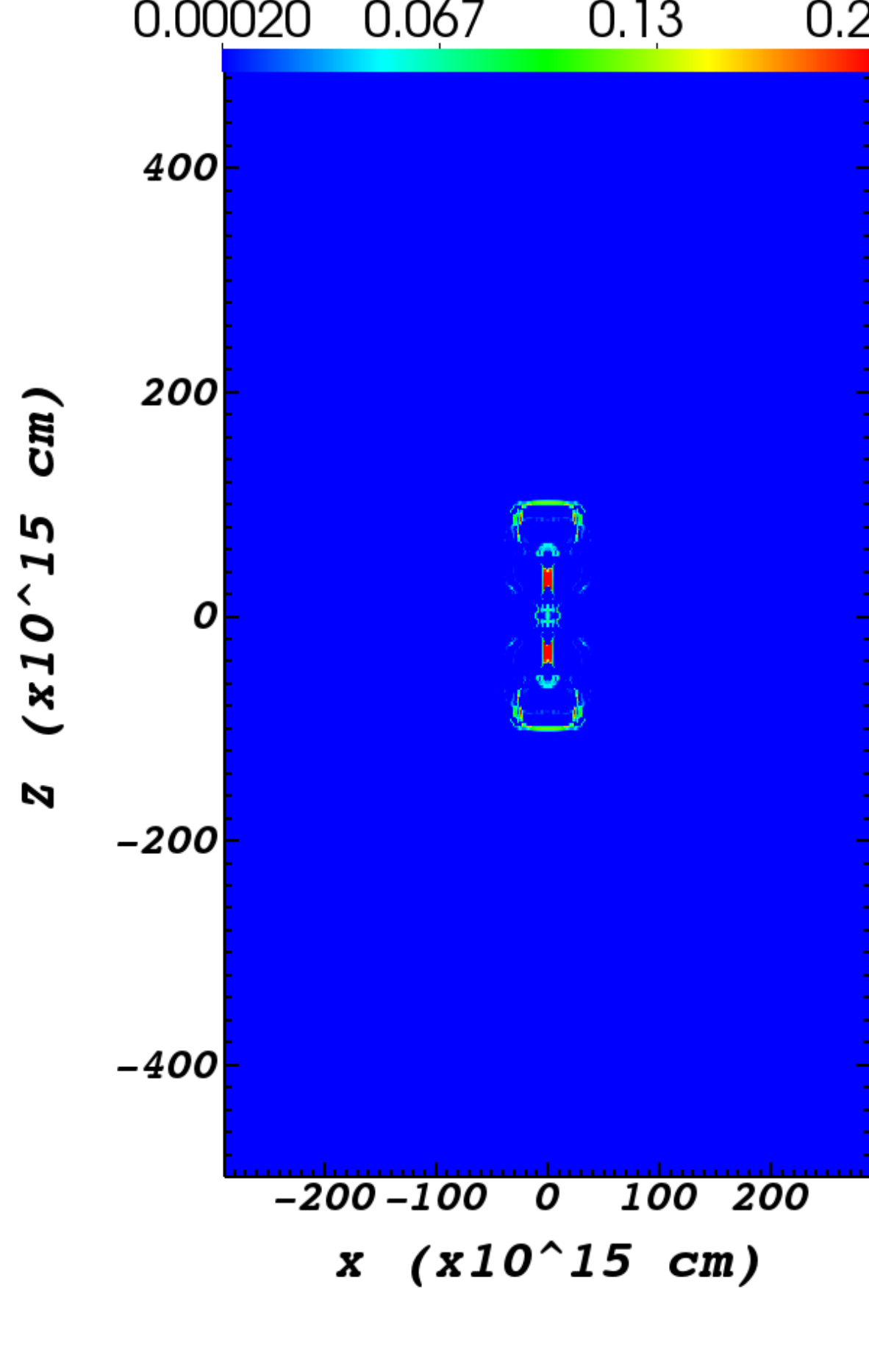}
\includegraphics[trim= 1.0cm 0.2cm 7.5cm 0.2cm,clip=true,width=0.33\textwidth]{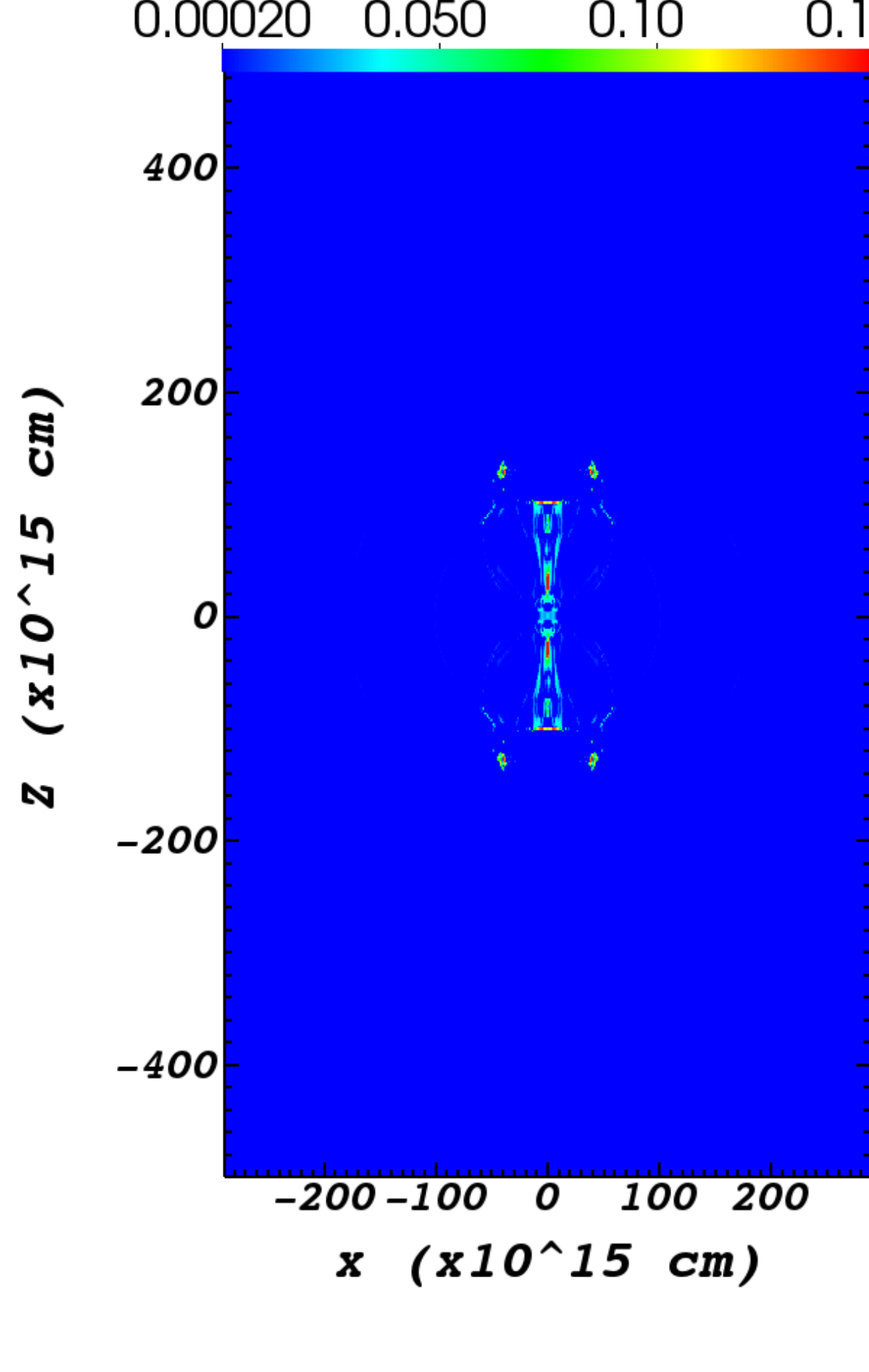}
\newline
\includegraphics[trim= 1.0cm 0.2cm 7.5cm 0.2cm,clip=true,width=0.33\textwidth]{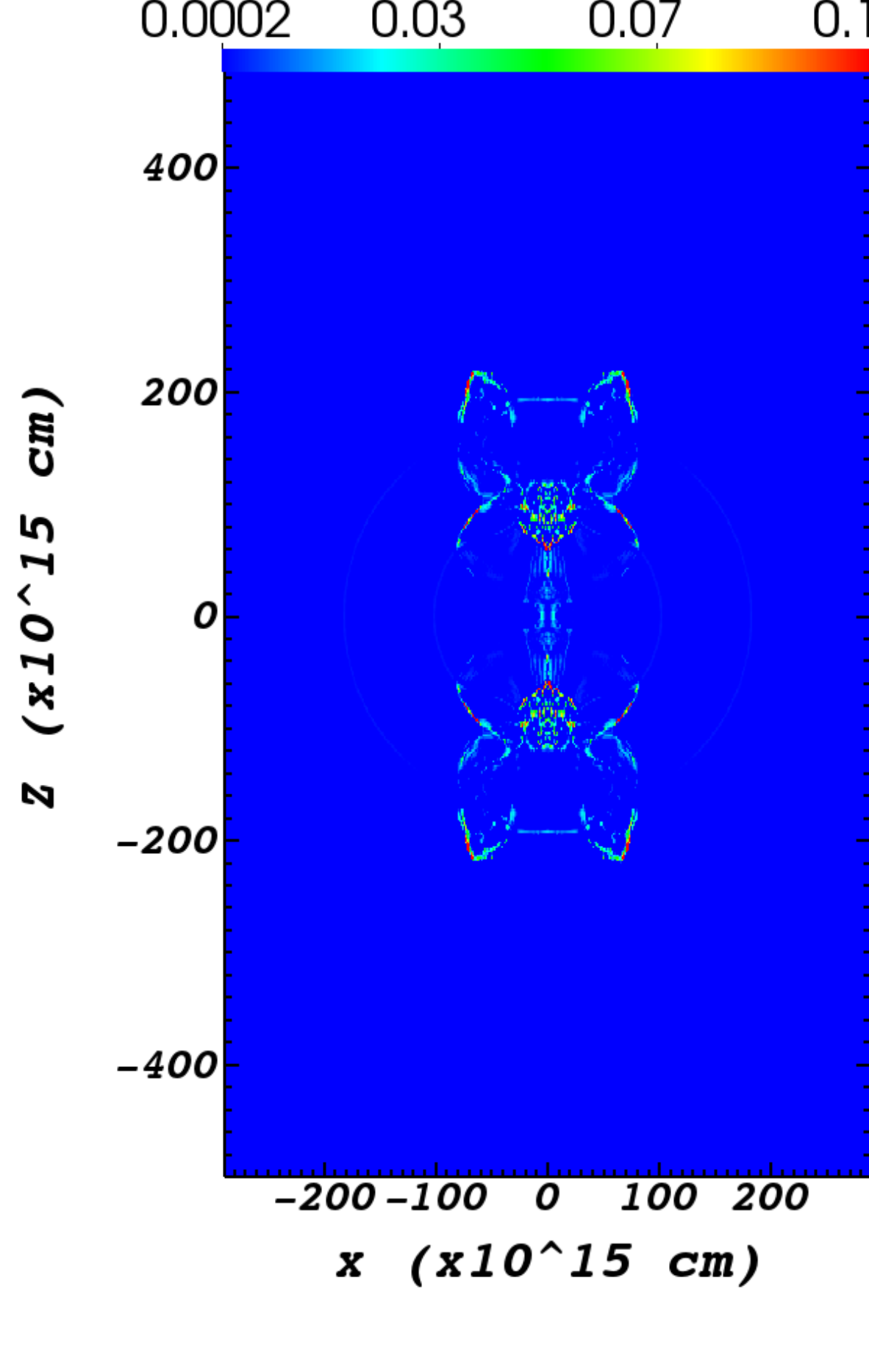}
\includegraphics[trim= 1.0cm 0.2cm 7.5cm 0.2cm,clip=true,width=0.33\textwidth]{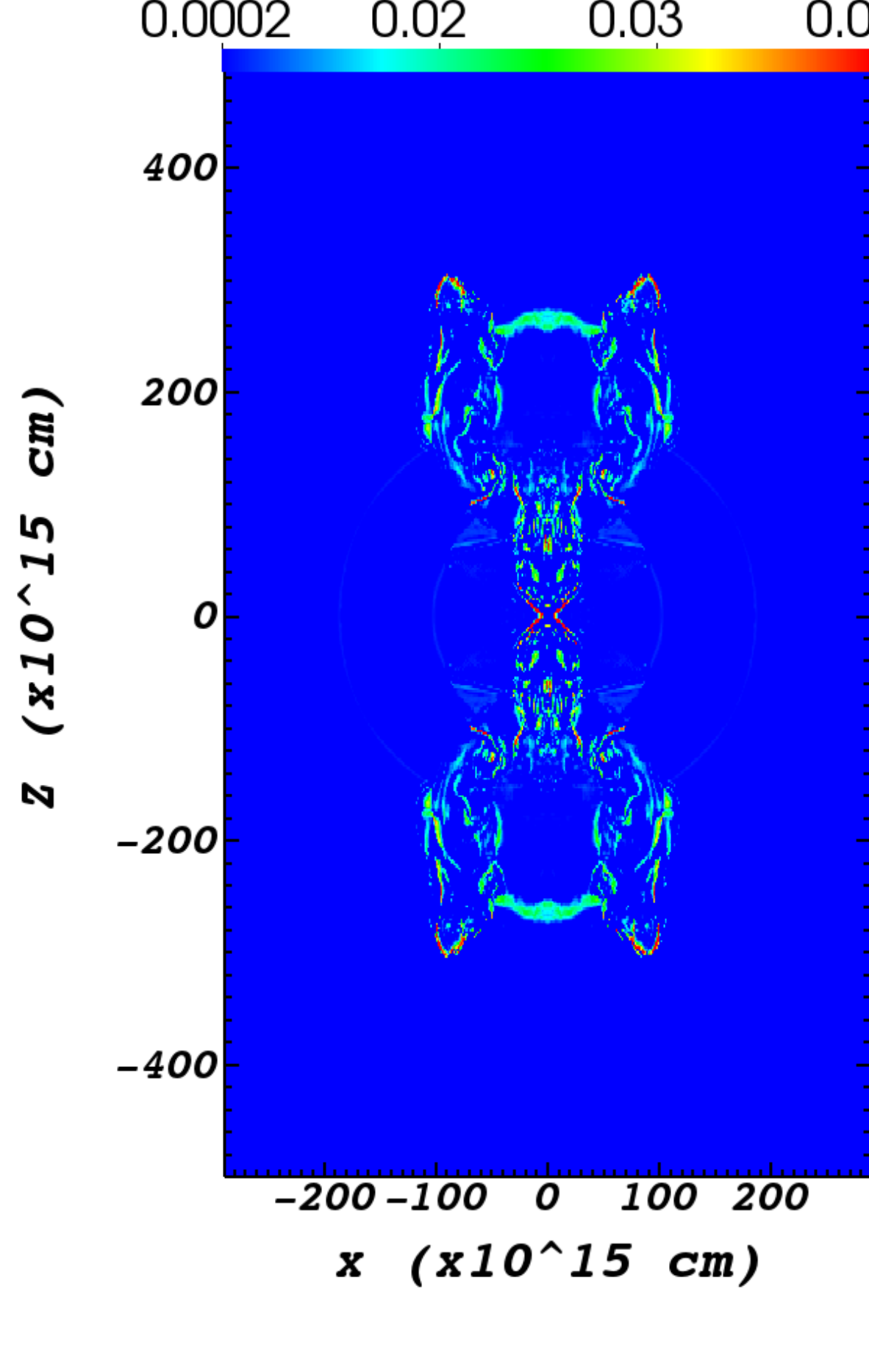}
\includegraphics[trim= 1.0cm 0.2cm 7.5cm 0.2cm,clip=true,width=0.33\textwidth]{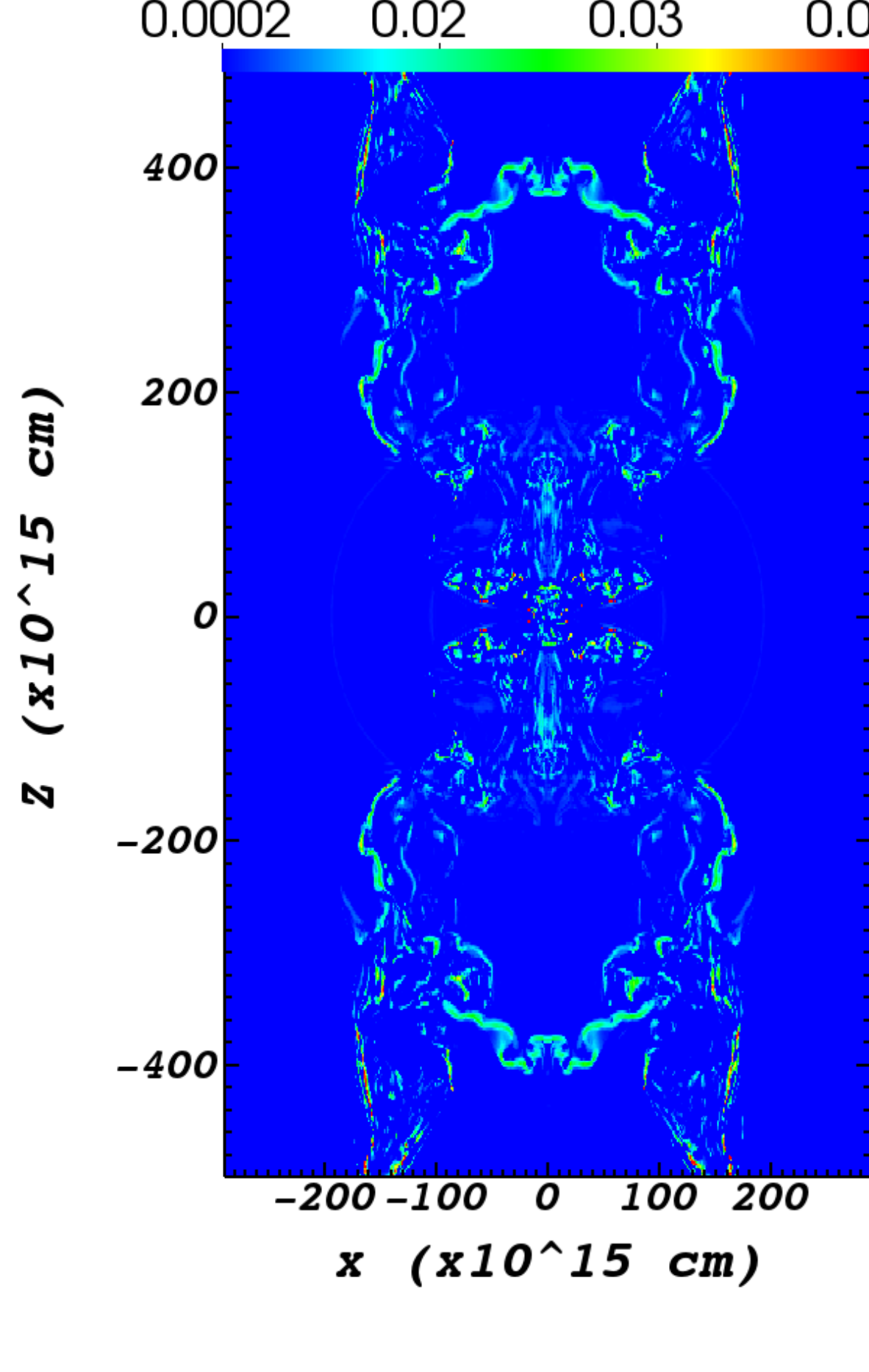}
 \caption{Maps of the quantity $\tau^{-1}_0$ as given by equation (\ref{eq:tau0}) and in units of $\yr^{-1}$, in regions that are  Rayleigh-Taylor unstable. Higher values means higher growth rates of the instabilities. The red color in the panels stands for (from upper left to lower right) , $0.2$, $0.2$, $0.15$, $0.1$, $0.05$, and $0.05 \yr^{-1}$, respectively.  } 
\label{fig:RT}
\end{figure}

 The interaction of the jets with the inner slow wind, and then the interaction of the expanding bipolar structure with the dense shell and outer slow wind are prone to Rayleigh-Taylor instability modes. The basic condition for the Rayleigh-Taylor instability to develop here is that the density gradient and the pressure gradient have opposite sense. Namely, if 
 $\overrightarrow {\nabla} \rho \cdot \overrightarrow{\nabla} P <0$, where $\rho$ is the density and $P$ is the pressure, then this region is Rayleigh-Taylor unstable.  
 In Fig. \ref{fig:RT} we present the quantity 
\begin{equation}
\tau_0^{-1} \equiv \rho^{-1} \left( - \overrightarrow {\nabla} \rho \cdot \overrightarrow{\nabla} P \right)^{1/2} \quad {\rm for} \quad 
\overrightarrow {\nabla} \rho \cdot \overrightarrow{\nabla} P <0, 
\label{eq:tau0}
\end{equation}
in unstable regions in the meridional plane and at six times as in earlier figures.  
The growth time of a Rayleigh-Taylor unstable mode of wavelength $\lambda$ is of the order of $\tau_{\rm RT} \simeq \tau_0 (\lambda/D_\rho)^{1/2}$, where $D_\rho \equiv \rho/ \vert \overrightarrow {\nabla} \rho \vert$.    

The unstable regions follow the bipolar structure. 
In the third panel at $t=70 \yr$ there are two unstable regions that look like `ears' on the front of each bipolar lobe (one at each side of the equatorial plane). These are the early development of the columns from instabilities. The `ears' and their development to columns are seen also in the last four panels of Figs. \ref{fig:dens_slice} - \ref{fig:vel_arrows}. An `ear' is the structure that forms behind a dense bullet that moves supersonically through the ambient gas.   

The fourth panel of Fig. \ref{fig:RT} at $t=120 \yr$ presents an interesting structure, in showing a large unstable region in the contact between the inner and outer bubble. Few red filaments in that region imply rapid growth rate of the instability. This region coincides with the inner radius of the dense shell, $r=10^{17} \cm$. 

\begin{figure}
 \begin{center}
\hskip -0.5 cm
 \includegraphics[trim= 1.0cm 0.2cm 0.0cm 0.2cm,clip=true,width=0.50\textwidth]{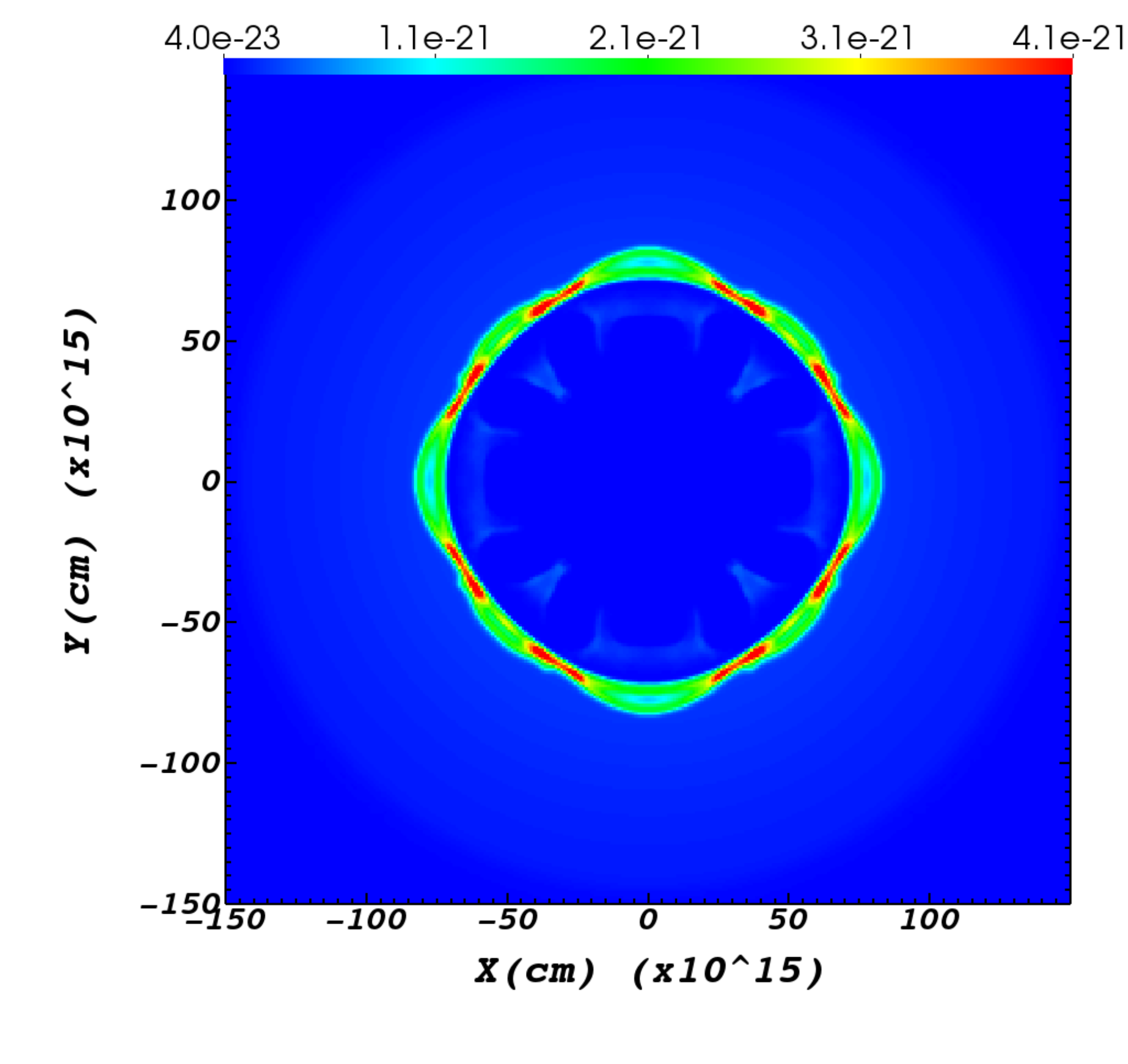}
\includegraphics[trim= 1.0cm 0.2cm 0.0cm 0.2cm,clip=true,width=0.50\textwidth]{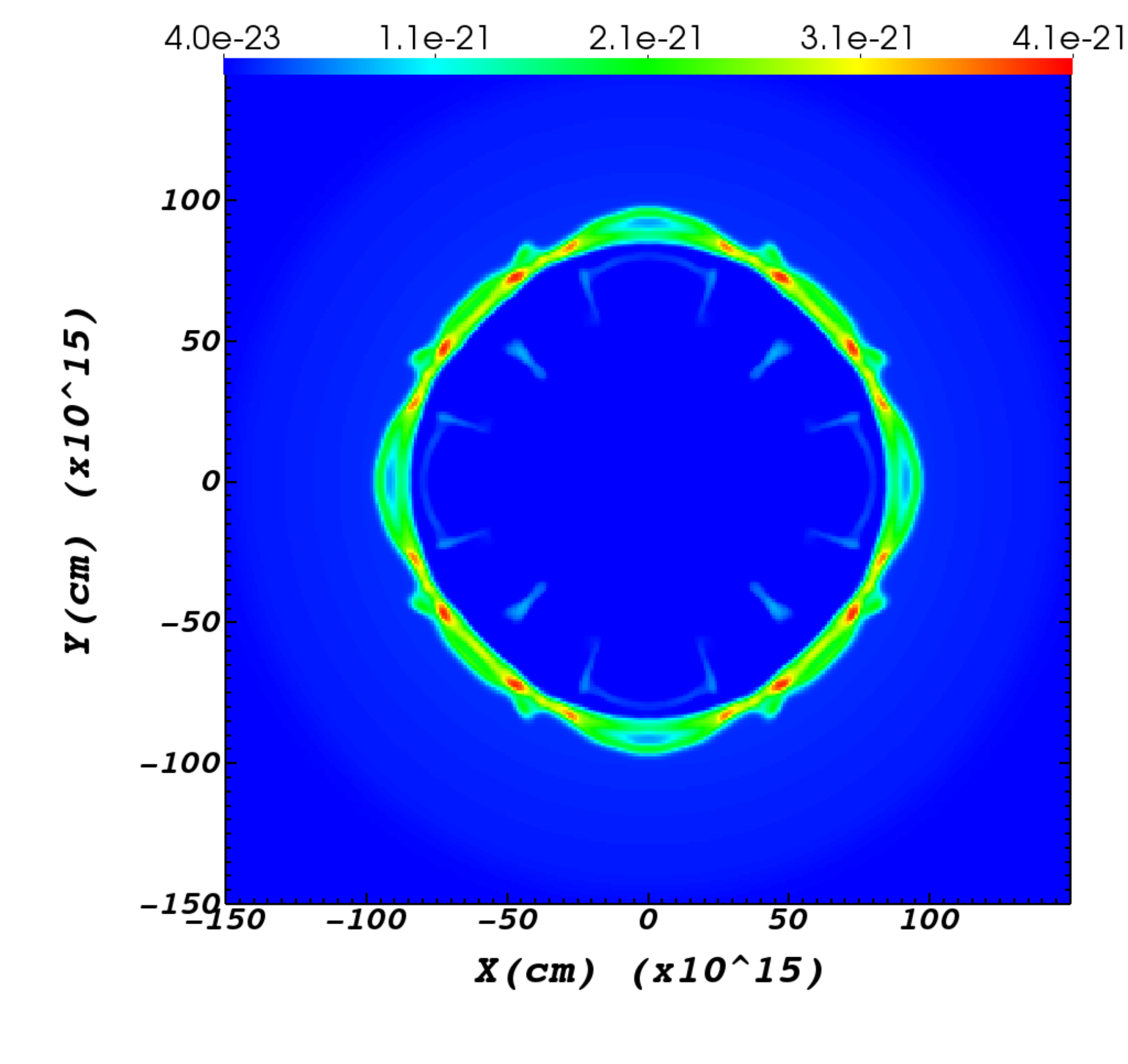}
 \vskip -0.3 cm
 \caption{The density maps in the $z=10^{17}\cm$ plane (parallel to the equatorial plane) at two times: $177 \yr$, and $273 \yr$, corresponding to the last two panels of Fig. \ref{fig:3D}. The density scale is given by the colour-bar in units of $\g \cm^{-3}$.}
  \label{fig:z=1e17}
 \end{center}
 \end{figure}

The physical properties of the flow determine where and how fast the instabilities develop. However, in our case it is the numerical grid, by its structure and resolution, that determines which are the most unstable modes and where exactly they develop the fastest. The Cartesian grid we use leads to the formation of 8 symmetrical columns, or four pairs of columns. In Fig. \ref{fig:z=1e17} we present the locations of the columns in the plane $z=10^{17} \cm$, i.e., parallel to the equatorial plane. The location of the columns are the high density spots that the red color represents in that figure. We suggest that in reality there are many more columns and that they are thinner, as there is no resolution limit. This implies that the bullets that we obtain (see Fig. \ref{fig:Scematic}) will be smaller, and will not survive for a long time. We might not observe them at late time.   

Let us further comment on our resolution. When the perturbations first develop, as we can see in Fig. \ref{fig:dens_slice} at $t= 44 \yr$, the width of the hight density fingers (4 long radial filaments in red) is about $2.5 \times 10^{15} \cm$ and the bullets  are resolved then by only 2.5 cells. This shows that the grid-cells determine the size of the perturbations that develop later to bullets. At the end of our simulation the bullets have a width of about $30 \times 10^{15} \cm$ and each one is resolved by about 30 cells. The bullets survive to the end of our simulation, although their structure is not smooth. They are shaped like a boomerang and the density is not constant.

One can define the `bullet crushing time' (e.g., \citealt{Jonesetal1996}) 
\begin{equation}
t_{\rm bc} = \frac{D \sqrt{\chi}}{v_{\rm rel}} \simeq 60
\left(\frac{D} {3 \times 10^{16} \cm} \right)^{1/2} 
\left(\frac{v_{\rm rel}} {500 \km \s^{-1}} \right)^{-1} 
\left(\frac{\chi} {10} \right)^{1/2}  \yr,
\label{eq:tbc}
\end{equation}
where $v_{\rm rel}$ is the relative velocity of the bullet and the ambient gas, $D$ is the size of the bullets, and $\chi$ is the ratio of the bullet density to the ambient density. We have substituted typical values at the end of our simulation. 
\cite{Jonesetal1996}, for example, show that after a time of several $t_{\rm bc}$ the bullet is highly distorted, although still exist. We have $t\simeq 5 t_{\rm bc}$ at the end of our simulation, and the bullets are distorted, but they are not `crushed' yet. It is possible that the numerical viscosity maintains the bullets intact for a longer time. A higher resolution simulation with lower numerical viscosity would form smaller bullets that live for a much shorter time. The result will be slower columns that might fit better the observations of Mz~3, as we discuss in section \ref{sec:summary}.

\section{DISCUSSION AND SUMMARY}
\label{sec:summary}

\begin{figure} 
\begin{center}
\vskip -0.7 cm
\includegraphics[trim= 1.5cm 16.0cm 2.0cm 3.0cm,clip=true,width=0.95\textwidth]{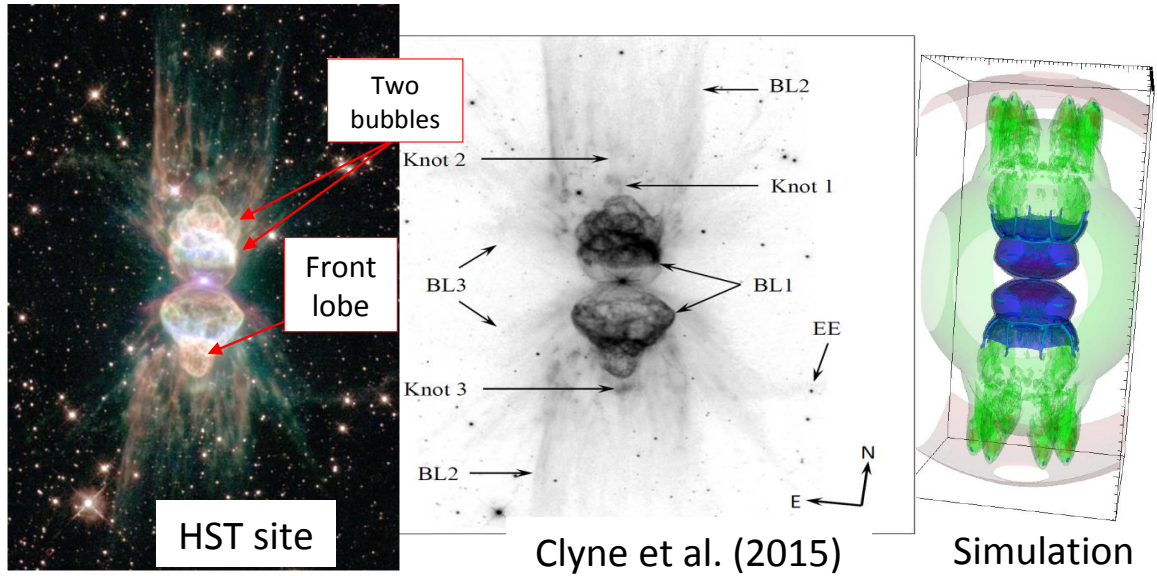}
\caption{Left panel: A false-color image of the PN Mz~3 by the Hubble Space Telescope. (Credit: NASA, ESA and The Hubble Heritage Team (STScI/AURA); Acknowledgment: Raghvendra Sahai and Bruce Balick.). Middle panel: An image of Mz~3 with marks from \cite{Clyneetal2015}. Right panel: Our numerical 3D density structure. }
 \label{fig:ant}
\end{center}
\end{figure}

The purpose of this study is to enrich the variety of morphological features that jets can form when they interact with the slow outflow from evolved giant stars. The rich variety of morphological structures of PNe and other nebulae around evolved stars motivate us to conduct these hydrodynamical simulations. Our motivation to conduct this study is the beautifully complicated structure of the PN Mz~3, the Ant nebula. In Fig. \ref{fig:ant} we present this PN. 

As with previous studies in this series, we set very simple initial conditions, i.e., a spherical slow ambient medium and one jet-launching episode. In the present study we set one dense shell within a slow wind from an AGB star, and injected two opposite jets in one episode that lasted $17 \yr$. In reality, the slow outflow is expected to be more complicated than we assume, and there can be several jet-launching episodes. For example, the jets do not need to  encounter a closed shell, but rather it is sufficient that each jet catches up and interacts with a dense slow polar cap with a half opening angle of $\ga 25^\circ$. Such caps can be ejected by an earlier slow bipolar outflow.
This will alleviate the non-detection of such spherical shells in PNe (e.g., \citealt{Sahaietal2011}).

This type of interaction forms a bipolar nebula with two prominent morphological features on each of the two sides of the equatorial plane, as we mark on Fig. \ref{fig:Scematic}: (1) A bipolar lobe that is composed of two bubbles, and (2) a columns-crown. 

In Figs. \ref{fig:3D} and \ref{fig:dens_slice} - \ref{fig:vel_arrows} we presented the evolution of the interaction. After the jets cease to exit, the bipolar structure that the jets inflated continues to move forward. The interaction with the dense shell splits each lobe into two bubbles, one touching the center and one further out.  

The interaction at its several stages has many regions that are Rayleigh-Taylor unstable. In Fig. \ref{fig:RT} the instability maps show these unstable regions that form the columns crown and the filaments and clumps inside the outer bubble.  

Let us compare the bipolar structure that we obtained in our simulation with the structure of Mz~3. We summarize the comparison in Table \ref{table:Compare}. 
Although we terminate the simulation at an age of about 300 years, the outer parts of the nebula, including the columns crown, have very high Mach numbers, $\simeq 10$, and hence experience a ballistic motion that preserves the shape of the nebula. We therefore can apply our results to the older PN Mz~3.
Each of the two lobes of Mz~3 is composed of two bubbles. However, these are not exactly the same as the two bubbles we obtained here. While in our simulation the outer bubble is larger than the inner one, in Mz~3 the outer one is smaller than the inner one. In a previous paper \citep{AkashiSoker2008a} we termed these outer small bubbles `front lobes'. In that paper we set different initial conditions, in particular there was no dense shell, and show how a jet can inflate a front lobe. 
Based on the earlier paper and the present study, we suggest one of the following two possibilities. (1) The exact setting of the dense ambient gas is more complicated, with some structure in between the setting we used in the two runs. (2) There were two jets-launching episodes in Mz~3, one episode that formed the two opposite crowns and one that formed the two front-lobes.
  We also note that a simple way to form two touching bubbles is to have two jets-launching episodes, one after the other. This requires no dense shell. 
\begin{table*}[t]
\small
 \centering \caption{Observed and simulated properties of Mz~3.}
 \begin{tabular}{c c c c} 
 \hline
 Property          &     Observed       &  Simulated           & Possible solutions for discrepancies\\
 [0.5ex]
 \hline \hline
General structure & Elongated           &Reproduced            &   \\
                  & Bipolar nebula      &                      &   \\             
  \hline
Columns-crown     & Marked BL~2         &Reproduced            &   \\
                  &in fig. \ref{fig:ant}&                      &   \\
  \hline
Opening of the    & Very small (almost  &  $25^\circ$ to the   & A more complicated initial  \\
crown             &parallel to axis)    &  symmetry axis       & ambient gas structure   \\                  
  \hline
Velocity of       & $\simeq 100 \km \s^{-1}$   &  $\simeq 500 \km \s^{-1}$   &  Slower jets\\  
columns           &    &     &  \\  
  \hline
Material near     & Not observed.       & Leftover from        &  Interaction closer to center \\  
equatorial plane  &                     & dense shell          &  followed by ballistic expansion\\  
  \hline
Front lobe        & clear on one        & Not reproduced       &  A later jet launching episode\\  
(Fig. \ref{fig:ant})& side                &                    &  as in \cite{AkashiSoker2008a} \\
  \hline
Two connected     & Outer bubble is     & Two bubbles, but     &  A somewhat different initial \\  
bubbles (Fig. \ref{fig:ant}) & smaller  & outer is larger      &  ambient medium structure \\                 
  \hline
Side rays         & Marked BL~3         & Not reproduced       &  A jet-launching episode \\  
                  &in fig. \ref{fig:ant}&                      &  of wide jets\\  
   \hline 
Width of columns  & Narrow              & Wider than observed  &  Simulate with a higher  \\  
                  &                     &                      &  resolution and a somewhat \\
                  &                     &                      &  different ambient structure \\ 
\hline 
Origin of columns &Extended zone        & A ring on the lobe.  &  Several short jets or a \\  
                  &on the lobe.         &                      &  longer one jets-episode  \\  
   \hline 
 X-ray emission   &  Observed inside    &  Simulation has high & Late jets and a fast wind  \\  
                  & the lobes           &  Temperature gas; not&  might improve the structural \\  
                  &                     &exactly same structure&  fitting \\
   \hline
\end{tabular}
\label{table:Compare} 
\newline
Comparison of the observed properties of Mz~3 with the results of the present simulation.
\end{table*}

Table \ref{table:Compare} list many properties of Mz~3 that we did not reproduce in the present simulation. There are two options to reconcile the discrepancies. The first is that our suggested jets-shell interaction is not the process that formed the columns crowns. The second one is that jets-shell interaction is the explanation to the columns-crowns, but there are several missing ingredients from our simulations. In the last column of Table \ref{table:Compare} we list these possible missing ingredients. The Table shows that we fail to reproduce most of the features of Mz~3. Future studies will have to examine other processes that can form the columns-crowns instead of jets, or else will reproduce with much more complicated simulations all properties of Mz~3, but leave jets as the explanation for the columns crown.

The distance to Mz~3 is uncertain, with different studies listing values in the range of about $1-3 \kpc$ (see discussion by \citealt{Clyneetal2015}). At a distance of $1.5 \kpc$ the width of the main nebula we obtained in our simulation  is about $20 ^{\prime \prime}$. This is compatible with the observed width of the main nebula (e.g., \citealt{Clyneetal2015}). \cite{SantanderGarciaetal2004} find the maximum velocity of the columns of the crown to be about $300 \km \s^{-1}$. The velocity of the columns that \cite{Clyneetal2015} observe (BL2 in their nomenclature), on the other hand, is only about $100 \km \s^{-1}$. We find that the velocity of the bullets in our simulation is about $500 \km \s^{-1}$, and the velocity of the column that trails each bullet is somewhat lower. This shows that we did not reproduce the structure of Mz~3 quantitatively, but only qualitatively, but not perfectly. To obtain a better match we will have to thoroughly study the parameter space.
  
 The two opposite columns crowns we have obtained here resemble the `crowns' in Mz~3, but they are not identical. Firstly, The `columns' in Mz~3 are much narrower than what we obtained here, and they better be termed filaments. Secondly, the `columns' in Mz~3 originate from an extended region on the lobes of Mz~3, while our columns originate from one ring on the outer bubble (Fig. \ref{fig:z=1e17}). 
  We suggest to reconcile these differences as follows. The reason we obtain thick columns is because of the limited resolution of the grid. A much finer grid (beyond our capabilities) might form thiner columns. As for the origin of the filaments. We used a short jets-launching episode that lasted only 17 years. It is possible that a longer episode, or several short episodes, might lead to a more extended crown. 

There are other features in Mz~3 that we do not reproduce. One example is the columns at very large angles, marked as BL3 by \cite{Clyneetal2015}. Another is the exact structure of the X-ray emission within the main nebula \citep{Kastneretal2003}. Although we do obtain high temperature gas that emits X-ray, it is not clear it will survive for a long time, and the exact structure is not as that in Mz~3. Although we do not reproduce the exact X-ray structure, the X-ray structure does support shaping by jets \citep{Kastneretal2003}
We attribute the more complicated observed nebular structure to two types of effects. The first is a much more complicated slow wind prior to the launching of jets, and the second is later jets-launching episodes. On top of these, the central star can blow a fast wind. Adding more features to the slow wind, and adding more jets-launching episodes make the parameter space much too large to follow. For that reason in our different studies we tend to study one type of interaction at a time.  
   
We summarize by stating that our main finding in the present study is that the type of Rayleigh-Taylor instability modes that develop in our simulation when jets and the lobes they inflate encounter circumstellar matter with large density gradients, larger than a wind with a constant mass loss rate, can account for the outward extending columns/filaments in Mz~3 and other nebulae. 

We thank Bruce Balick and an anonymous referee for enlightening and for useful comments.  We acknowledge support from the Israel Science Foundation and a grant from the Asher Space Research Institute at the Technion.


\label{lastpage}
\end{document}